\documentclass[twocolumn,showpacs,preprintnumbers,amsmath,amssymb,amsfonts,
nofootinbib,
 aps,
 prb,
floatfix,
 lengthcheck,%
]{revtex4-2}
\usepackage[english]{babel}
\usepackage[utf8]{inputenc} 
\usepackage{SIunits}
\usepackage{hyperref}
\usepackage{graphicx}
\usepackage{color}

\begin{document}
\title{Microscopic theory of photo-assisted electronic transport in \\normal-metal/BCS-superconductor junctions}
\author{Bruno Bertin-Johannet}
\email{bruno.bertin@cpt.univ-mrs.fr}
\author{Jérôme Rech}
\author{Thibaut Jonckheere}
\author{Benoît Grémaud}
\author{Laurent Raymond}
\author{Thierry Martin}
\affiliation{Aix Marseille Univ, Universit\'e de Toulon, CNRS, CPT, Marseille, France}

\begin{abstract}
We investigate  photo assisted electronic transport in a normal-metal/BCS-superconductor junction with a  microscopic Hamiltonian approach, for several types of periodic voltage drives applied on the normal-metal side. The time-dependent current and the photo-assisted noise are computed to all orders of the tunneling Hamiltonian using a Keldysh-Nambu-Floquet approach. An excess noise analysis allows one to determine to what extent pure electronic excitations with a small number of electrons per period can be generated by the different drives. When the superconducting gap is small compared to the drive frequency, the junction behaves like a normal-metal junction and minimal excess noise is reached for Lorentzian voltage drives carrying an integer charge (levitons). In the opposite regime of a large-gap, the excess noise vanishes for half-quantized levitons, giving rise to the perfect transmission of a Cooper pair on the superconducting side. This microscopic approach also allows us to address the intermediate regime, when the drive frequency is comparable to the gap, allowing us to study the non-trivial interplay between Andreev reflection and quasiparticle-transfer processes. Our analysis also shows the appearance of Tien-Gordon-type relations connecting the current and noise in the AC-driven junction to their DC counterpart, which we investigate in details. Finally, the possibility to build a  reliable on-demand source of Cooper pairs with this setup is examined using realistic experimental parameters.
\end{abstract}

\maketitle

\section{Introduction}\label{Intro}
	Electron quantum optics (EQO) aims at describing and manipulating single electronic excitations in condensed matter systems. This is achieved by adapting scenarios of quantum optics where, for instance, single photons are sent on a beam splitter. This includes the Hanbury-Brown and Twiss experiment~\cite{brown1956a} where the intensity correlations from coherent photons at the output are observed. Alternatively, in the Hong-Ou Mandel setup,~\cite{hong1987a} photons collide at the location of the beam splitter and correlations are measured at the output. In condensed matter settings, electron wave guides can be achieved with a two dimensional electron gas (2DEG), while a quantum point contact (QPC) mimics the beam splitter. However, electrons differ from photons as they are charged particles and bear fermionic statistics. This means, in particular, that they interact strongly with their neighboring electromagnetic environment and are always accompanied by a Fermi sea.

In recent decades, the combination of theoretical~\cite{levitov1996a} and experimental~\cite{dubois2013b} efforts, boosted by advances in fabrication techniques, has provided EQO with a strong foothold. In particular, a range of results can be interpreted in terms of a Fermi liquid picture. Concerning single electron sources, special interest has been devoted to trains of quantized Lorentzian pulses~\cite{levitov1996a} dubbed levitons. Levitons consist of ``pure'' single electron excitation~\cite{keeling2006a}, i.e., devoid of unwanted electron-hole pairs. When a combination of AC and DC bias is applied at the entry ports of a QPC, the measurement of the output excess noise (with respect to. the proper reference situation with only an applied DC bias) allows the detection of these spurious electron-hole excitations. Yet, the Fermi liquid picture has to be revisited when decoherence effects~\cite{jonckheere2012a} or embedded correlations (as in the fractional quantum Hall effect~\cite{rech2017a}) operate, requiring an adapted formalism and yielding new effects such as charge fractionalization~\cite{wahl2014a} or leviton crystallization~\cite{ronetti2018a}.

While so far mostly Coulomb repulsion has been effectively included in such scenarios, other types of correlations also deserve consideration. Indeed, electron waveguides can be connected to superconducting leads, opening the way to hitherto unexplored EQO effects, such as electron (respectively hole) conversion into Bogoliubov quasiparticles~\cite{blonder1982a} above (respectively below) the gap or Andreev reflection~\cite{andreev1964a} (AR) of electrons or holes inside the gap~\cite{acciai2019a}.
This is precisely the goal of this study: we address, using a microscopic model, how electron pulses generated by a voltage drive on a normal-metal behave at a tunnel junction with a superconductor, note that it differs from the setup studied in Ref.~[\onlinecite{acciai2019a}] involving two superconductors, where only quasiparticle transfer is considered within a perturbative scheme. 

This system was discussed earlier by Belzig \textit{et al.}~~\cite{belzig2016a} through full counting statistics of the electric current in the context of circuit theory~\cite{nazarov1999a}. In this study, they considered the zero temperature limit and focused on the two limiting regimes where the drive frequency is either much larger or much smaller than the gap of the superconductor. In the former situation, they found that the excess noise (XN) is suppressed for integer charge carrying levitons,  in accordance with a normal-metal junction. In the latter one, where transport is dominated by Andreev reflection, they found excess noise suppression also for levitons carrying half-integer charge. The effect of a finite temperature or the fate of the junction in the intermediate regime between these two limiting cases remained largely unexplored, and this is the gap we intend to bridge here.

In this work, we develop a microscopic Hamiltonian model of the junction, allowing us to compute the average current as well as the period-averaged noise at all orders in the tunneling constant using Green's functions in the Keldysh formalism. This enables access to all regimes for the relevant parameters, providing analytical derivations when possible. Not only can we describe the junction over the whole range of driving frequency (smaller than, comparable to or above the superconducting gap) but our approach also allows us to exactly account for finite temperature. The formalism we use is quite versatile, and allows us to tackle any type of periodic drive (we typically restrict ourselves to sinusoidal, square, and Lorentzian drives). It heavily relies on Floquet theory~\cite{pedersen1998a}, where as a consequence of the applied AC drive, electrons can absorb or emit photons leading to the formation of side bands in energy, or Floquet channels. These are populated with probabilities directly connected to the Fourier decomposition of the exponentiated drive, or so-called Floquet weights. 

We start by considering the regime of vanishingly small gap, where results naturally coincide with those of a $N-N$ junction, with excess noise suppression obtained when the applied voltage is a train of levitons with integer charge~\cite{dubois2013a}. We then focus on the opposite regime of an infinite gap, where the results can now be cast in a form similar to the $N-N$ case, but with Andreev reflection replacing electron transmission, and the excess noise gets suppressed when the applied voltage is a train of levitons with half-integer charge~\cite{belzig2016a}. Within our framework, we are able to provide a microscopic argument explaining why the XN vanishes for these specifically tailored pulses. We also describe the crossover regime in which the driving frequency is comparable to the superconducting gap. There, we derive an exact expression for the average current, and also provide a detailed analysis of the excess noise, relying first on an analytic perturbative expansion at low order in the tunneling constant, before solving numerically the full problem at all orders.  Our results can be interpreted in terms of Floquet transport channels, uncovering the importance of ``effective gaps'' corresponding to the superconducting gap as seen from a given Floquet channel.

Our analytic derivation also allows us to establish that current and noise satisfy Tien-Gordon-type relations,~\cite{tien1963a} i.e. that there is a profound connection between AC and DC driven behaviors. While this is always satisfied for the average current, which can be viewed as a weighted sum of independent contributions from each Floquet channel, it is only valid for the noise in the limiting regimes. We propose an interpretation for these results in terms of the relevant physical processes at play and the interference effects expected to occur between different Floquet channels.

Finally we consider the possibility to use the driven $N-S$ junction as a source of Cooper pairs.  Building on our understanding of the various regimes, we analyze the effects of a finite temperature along with the departure from perfect transmission, and show that there exists a set of experimentally accessible parameters for which the $N-S$ junction driven by an appropriately tuned periodic Lorentzian drive operates as a reliable source of Cooper pairs with a properly quantized average transmitted charge as well as minimal excess noise. 

The paper is organized as follows. In Sec.~\ref{ModelSec} we introduce the theoretical framework for tunneling through the junction in the presence of a periodic drive. In Sec.~\ref{sec:limits}, we analytically recover known results for the two limiting regimes~\cite{belzig2016a} providing an interpretation for the relevant signatures within our formalism. Sec.~\ref{InterSec} then focuses on the intermediate regime where the drive frequency is comparable to the gap. In Sec.~\ref{sec:TGlike}, we investigate in more details the Tien-Gordon-type relations we uncover for the current and noise.
A scheme to design an on-demand source of Cooper pairs realizable experimentally is discussed in Sec.~\ref{OnDemand}. We conclude in Sec.~\ref{Conclu}. Some additional technical aspects are presented in the Appendices. We adopt units in which $\hbar=k_B=1$ and the electronic charge is $e<0$. The temperature of the system is denoted $\theta$ and $\beta$ corresponds to the inverse temperature, i.e  $\beta^{-1}=k_B \theta$.

\section{Model}\label{ModelSec}

\subsection{Hamiltonian approach}
We adopt a similar approach to the one developed by Cuevas \textit{et al.}~\cite{cuevas1999a} for junctions involving superconductors in which the BdG equations are discretized. The left and right leads are described at equilibrium by the following Hamiltonians
\begin{equation}
	\begin{aligned}
		H_{\text{L}}&=H_{0,\text{L}}\\
		H_{\text{R}}&=H_{0,\text{R}}+\Delta\sum_i\left(c_{i,R,\downarrow}^\dagger c_{i,R,\uparrow}^\dagger +c_{i,R,\downarrow}c_{i,R,\uparrow}\right),
	\end{aligned}
\end{equation}
where $H_0$ is the kinetic part of the Hamiltonian, $i$ labels the various sites of these leads, $\Delta$ is the superconducting gap and the chemical potential is set to zero in both electrodes.  Here $c_{i,L,\sigma}$ (respectively $c_{i,R,\sigma}$) is the electron annihilation operator at site $i$ on the left (respectively right), with spin $\sigma$.
\begin{figure}
	\centering
	\includegraphics[width=0.4\textwidth]{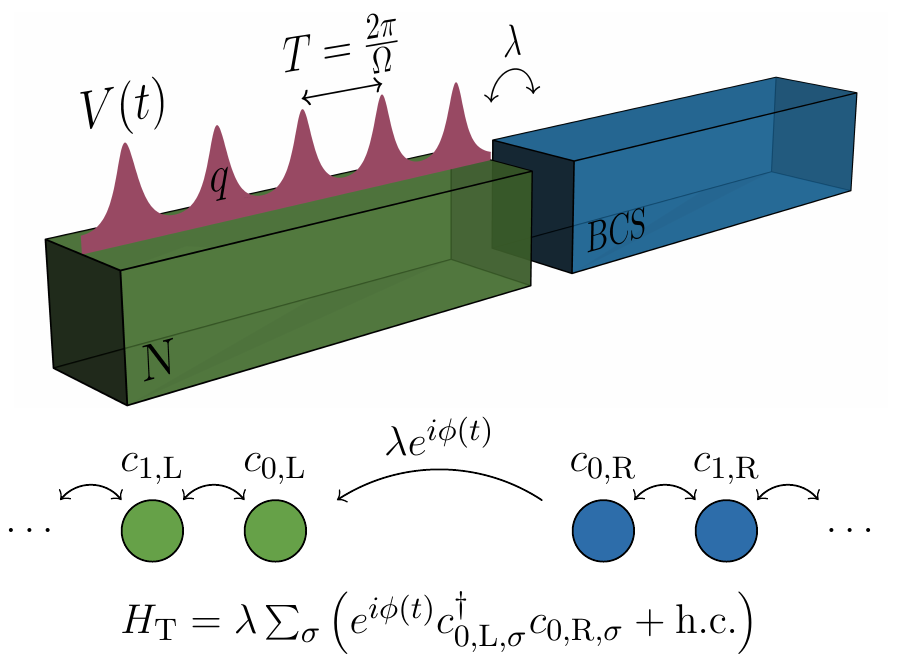}
	\caption{Simplified drawing of the junction considered. Top drawing represents both metals separated by a junction by tunnel coupling $\lambda$, the normal-metal is driven by a time dependent voltage drive $V(t)$ of period $T$. The bottom drawing represents the tight binding model describing the junction in the gauge where the time dependence is in the tunnel coupling.}
	\label{Setup}
\end{figure}
By convention, we consider that the coupling between the two leads occurs at sites $0$ (see Fig.~\ref{Setup}). One then defines the Nambu spinors
\begin{equation}
	\psi_L^\dagger=\begin{pmatrix}
		c_{0,L,\uparrow}^\dagger & c_{0,L,\downarrow}
	\end{pmatrix}\, ,\qquad \psi_R^\dagger=\begin{pmatrix}
		c_{0,R,\uparrow}^\dagger & c_{0,R,\downarrow}
	\end{pmatrix}
\end{equation}
allowing us to write the tunneling Hamiltonian between the leads as
\begin{equation}\label{CurrentHamApp}
	H_\text{Tun}= \psi_L^\dagger W_{LR} \psi_{R}+\text{H.c.}  \, .
\end{equation}
The total Hamiltonian therefore reads as
\begin{equation}
H=H_{\text{L}}+H_{\text{R}}+H_{\text{Tun}}
\end{equation}

The tunnel matrix between the coupled sites of the left and right leads of the junction is defined as
\begin{equation}
	W_{LR}=\lambda\sigma_z e^{i \sigma_z \phi(t)}\, ,
\end{equation}
with $\sigma_z$ the Pauli matrix in Nambu space, and $\lambda$ the tunneling amplitude. The phase
$\phi(t) = e\int_{-\infty}^{t} dt'\, V(t')$ is the time-dependent phase difference between the leads which accounts for the drive $V(t)$ applied on the left lead. Note that $W_{LR}^\dagger=W_{RL}$.

The current flowing from the left (normal) lead is obtained from the current operator defined as
\begin{equation}\label{Current}
	I_L(t)=ie\big[\psi_L(t)\sigma_zW_{LR}(t) \psi_R^\dagger(t)-\text{H.c.}\big]\, .
\end{equation}

\subsection{The average current and noise}
The first two moments of the current operator are computed in the framework of Keldysh theory~\cite{keldysh1965a}. We perform the time ordering on Keldysh contour and our convention is to use
\begin{equation}
	G^{+-}_{jj'}(t,t')=i\left\langle\psi_{j'}^\dagger(t')\otimes\psi_j(t)\right\rangle,
\end{equation}
where $G_{jj'}^{+-}$  is the Green function dressed by the tunneling Hamiltonian and $j$, $j'$ are lead indices.

It follows that the average current is expressed as a Nambu trace,
\begin{equation}
	\left\langle I_L(t)\right\rangle=e\text{Tr}_{\text{N}}\text{Re}\left[\sigma_zW_{LR}(t)G^{+-}_{RL}(t,t)\right]\, ,
\end{equation}
where $\text{Re}$ denotes the real part.

We are also interested in the mean deviation from the average current so we use the real time zero-frequency irreducible noise correlator, defined as
\begin{align}
	S_{LL}(t)=\int_{-\infty}^{+\infty}\mathrm{d}t'& \left[ I_L \left(t+t' \right) I_L \left( t \right) \right. \nonumber \\
	 & \left. \quad - \left\langle I_L \left( t+t' \right) \right\rangle \left\langle I_L \left( t \right) \right\rangle\right] \, .
\end{align}
Using Wick theorem, its average value becomes
\begin{widetext}
		\begin{align}
			\left\langle S_{LL}(t)\right\rangle=-e^2\int_{-\infty}^{+\infty}\mathrm{d}t'\text{Tr}_{\text{N}}\Big\{ & 2\text{Re}\left[\sigma_z W_{LR}(t) G_{RL}^{-+}(t,t')\sigma_z W_{LR}(t')G_{RL}^{+-}(t',t)\right]     \nonumber    \\
			-                                                                                                      & \sigma_z W_{RL} (t) G_{RR}^{-+}(t,t')\sigma_z W_{LR}(t')G_{LL}^{+-}(t',t) -\sigma_z W_{LR}(t) G_{LL}^{-+}(t,t')\sigma_z W_{RL}(t')G_{RR}^{+-}(t',t)
			\Big\}\, ,
			\label{NoiseTime}
		\end{align}
\end{widetext}
where we introduced $-+$ time ordered Green function
\begin{equation}
	G^{-+}_{jj'}(t,t')= -i \left\langle \psi_j(t)  \psi_{j'}^\dagger(t') \right\rangle .
\end{equation}
Note that the matrices entering this expression for the noise are all written in Nambu space.

\subsection{Voltage drive and Floquet theory} \label{sec:floquet}
\begin{figure*}
	\centering
	\includegraphics[width=\textwidth]{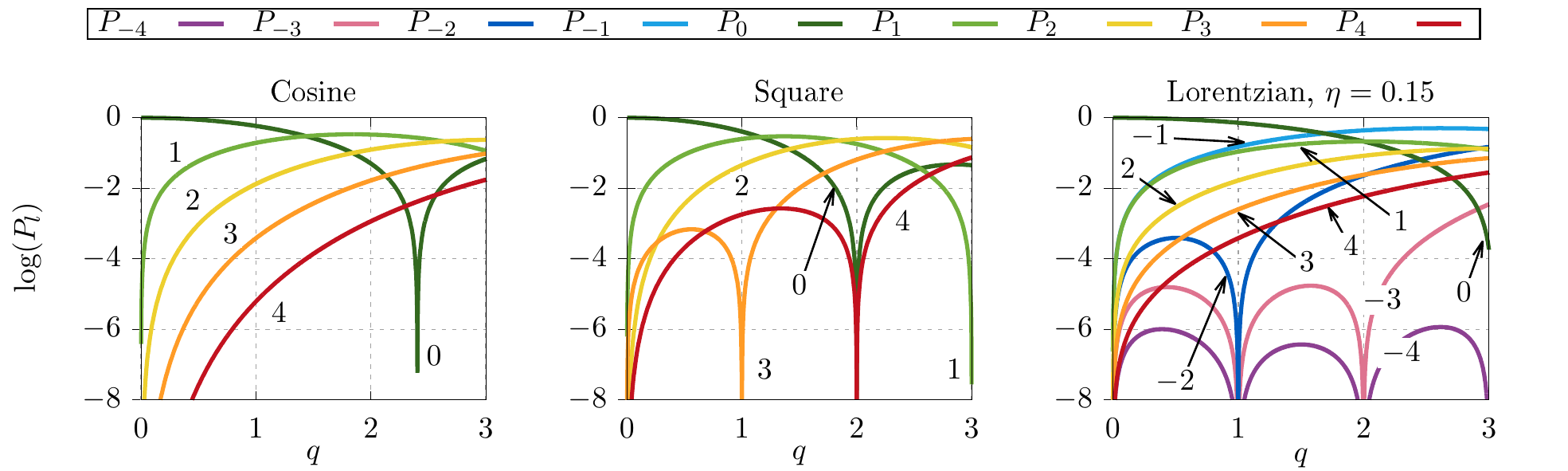}
	\caption{The logarithm of the Floquet weights of the first few channels, as a function of the injected charge per period $q$, for the three different drives, cosine, square and Lorentzian train of pulses (with $\eta=0.15$). The index of the coefficient corresponding to each curve is written on the graph. Note that $P_{-l} = P_{l}$ for the cosine and square drives.}
	\label{Floquet_Coefs}
\end{figure*}

We consider that a periodic drive of frequency $\Omega$
is applied on the normal side of the junction,
with the goal of injecting up to a few electrons per period. We are
particularly interested in the so-called levitons, which consist of a periodic train
of Lorentzian pulses, as they are able to excite an integer number of electrons without any other perturbation to the Fermi sea~\cite{lesovik1994a,dubois2013b}. For the sake of comparison, we also consider the case of a periodic cosine voltage, and of a periodic square voltage.

The periodic voltage can always be written as the combination of a DC and an AC part:
\begin{equation}
	V(t) = V_{\text{DC}} + V_{\text{AC}}(t)
\end{equation}
where $V_{\text{DC}}$ is time-independent, and $V_{\text{AC}}(t)$ averages to zero on one period
$T=2\pi / \Omega$ of the periodic drive. The DC part of the voltage determines the
injected charge per period. We define this important quantity as
\begin{equation}
	q = \frac{e}{2 \pi} \int_0^T \mathrm{d}t\, V(t) =  \frac{e V_{\text{DC}}}{\Omega} .
\end{equation}
Note that the drive affects both spin species in the same way, so an injected charge
of $q=1$ for example corresponds to a spin up electron and a
spin down electron injected per period.

In practice, the $DC$ component of the drive is actually fully taken into account by shifting the Fermi energy of the normal-metal by $eV_{\text{DC}}$, such that one is left dealing only with the $AC$ part of the drive. As the voltage appears in the Hamiltonian as exp$\left[ i e \int_{-\infty}^{t} dt' V \left( t' \right) \right]$,
the AC part of the voltage is best described by introducing the Fourier coefficients $p_l$
defined as:
\begin{equation}
	\label{p_lFormula}
	\mbox{exp}\left[ -i e \int_{-\infty}^{t} \!\!\! dt' \; V_{ac} \left( t'\right) \right]
	= \sum_l p_{l} \; e^{-i l \Omega t}\,.
\end{equation}
By doing so, we use Floquet theory~\cite{tien1963a,shirley1965a,pedersen1998a,moskalets2002,rech2017a} in which the total Hamiltonian is separated into an infinite number of \textit{independent} harmonics in Fourier space. The Floquet theory goes beyond this simple Fourier decomposition. Indeed, it states that as a consequence of the AC drive, the electrons can gain or lose energy quanta leading to the formation of side bands. The Floquet weight $P_l = \left\lvert p_l\right\rvert^2$ therefore corresponds to the probability for an incoming electron to absorb $l$ photons of energy $\Omega$. The voltage biased lead is then better described as a Floquet state, a superposition of Fermi seas, which we now refer to as ``Floquet channels'', with shifted chemical potential $\mu \to \mu + eV_{\text{DC}}+ l \Omega$ and an intensity given by the corresponding Floquet weight $P_l$. 

For a time-dependent system defined by Eq.~\eqref{CurrentHamApp}, one can show that the advanced/retarded self-energy entering the Dyson equation simply reads as $\Sigma(t,t')=\delta(t-t')W_{LR}(t)$ (see Appendix~\ref{DysonApp}). Following the formalism presented in Appendix~\ref{FourierApp},  this self energy becomes an infinite matrix $\hat{\Sigma}(\omega)$ expressed in the enlarged Nambu-harmonics space. More precisely, the matrix elements of this self-energy in harmonics space can be described by the following $2\times2$ block structure in Nambu
\begin{equation}
	\Sigma_{mn}(\omega)=\lambda\begin{pmatrix}
		p_{n-m} & 0        \\
		0       & -p^*_{m-n}
	\end{pmatrix}
\end{equation}
which turns out to be independent of $\omega$.
To emphasize the scaling of the self-energy in the tunneling amplitude $\lambda$,
we define the matrix $\hat{\mathcal{P}}$ as:
\begin{equation}
	\hat{\Sigma} = \lambda \, \hat{\mathcal{P}}   ,
	\label{Pcaldef}
\end{equation}
which we will use in the equations below for the current and the noise.

The expressions of the $p_l$ coefficients are given in Appendix~\ref{FloquetApp} for the three
different voltage drives that we consider:~\cite{rech2017a} the cosine drive, the square drive and a train of Lorentzian pulses. This latter choice involves an
additional parameter $\eta$, which corresponds to the ratio of the width of the Lorentzian shape divided by the period of the drive (in what follows, we consider values
of $\eta$ in the range 0.1-0.15, corresponding to experimentally accessible narrow Lorentzian pulses). As we consider voltages for which the amplitude of the AC component is directly related to the DC component, the $p_l$ coefficients only depend on the charge injected per period $q$.

The logarithms of the $P_l$ are represented in Fig.~\ref{Floquet_Coefs} for the first few values of the index $l$ and the various drives considered. We point out that, for small values of $q$, only the Floquet weights with the lowest $|l|$ contribute. When $q$ increases, the higher values of $|l|$ start to have a non-negligible contribution. One sees that, for Lorentzian pulses (right panel of Fig.~\ref{Floquet_Coefs}), $P_l\neq P_{-l}$ which is not the case for other drives. For all the drives, we remark that there are some parameters for which $P_l$ vanish. In the case of a cosine drive (left panel of Fig.~\ref{Floquet_Coefs}), this corresponds to the zeros of the Bessel functions of the first kind. For the square drive (middle panel of Fig.~\ref{Floquet_Coefs}), the Floquet weights vanish for $l-q=2n$ with $n$ integer, and those of the Lorentzian pulses  (right panel of Fig.~\ref{Floquet_Coefs}) vanish for integer values of $q$ which obey $q<-l$. We also notice that for the cosine drive, high index Floquet weights are well separated in magnitude which is not the case for the square drive.

In all generality, the current and the noise are complex time-dependent objects and we are instead primarily interested in their average value over one period of the drive. The time-averaged current can then be written as the energy integral of a Nambu-harmonics trace, namely
\begin{align}
\overline{\left\langle I_L\right\rangle} & =e \int_{-T/2}^{T/2} \frac{\mathrm{d}t}{T}\, \text{Tr}_{\text{N}} \text{Re} \left[\sigma_z W_{LR}(t)G_{RL}^{+-}(t,t)\right]                 \nonumber   \\
& =e \lambda \int_{-\Omega/2}^{\Omega/2}\frac{\mathrm{d}\omega}{2\pi}\text{Tr}_{\text{NH}} \text{Re} \left[ \hat{\sigma}_z  \hat{\mathcal{P}}^\dagger  \hat{G}_{LR}^{+-}(\omega)\right]\, .
		\label{Current_Fourier_global}
\end{align}
where $ \hat{\sigma}_j$ is a tensor product of the usual Nambu matrix $\sigma_j$ with the identity matrix in harmonics space.

A similar computation can be performed for the noise yielding the following expression for the zero-frequency period averaged noise (PAN)
\begin{widetext}
	\begin{align}
			\overline{\left\langle S_{LL}\right\rangle}\equiv \int_{-T/2}^{T/2}\frac{\mathrm{d}t}{T}\, \left\langle S_{LL}(t)\right\rangle & =-e^2\lambda^2\int_{-\Omega/2}^{\Omega/2}\frac{\mathrm{d}\omega}{2\pi}\text{Tr}_{\text{NH}}\Big\{2\text{Re}\left[\hat{\sigma}_z \hat{\mathcal{P}}^\dagger \hat{G}_{RL}^{-+}(\omega)\hat{\sigma}_z \hat{\mathcal{P}}^\dagger \hat{G}_{RL}^{+-}(\omega)\right] \nonumber \\
			                                                                                                                                             & \qquad\qquad\qquad - \hat{\sigma}_z \hat{\mathcal{P}} \hat{G}_{RR}^{-+}(\omega)  \hat{\sigma}_z \hat{\mathcal{P}}^\dagger \hat{G}_{LL}^{+-}(\omega)   -  \hat{\sigma}_z \hat{\mathcal{P}}^\dagger  \hat{G}_{LL}^{-+}(\omega) \hat{\sigma}_z \hat{\mathcal{P}}  \hat{G}_{RR}^{+-}(\omega)
			\Big\}\, .
			\label{Noise_Fourier_global}
		\end{align}
\end{widetext}

These expressions for the current and the noise can readily be evaluated once the dressed
Green functions  $\hat{G}_{ij}^{+-}$ and $\hat{G}_{ij}^{-+}$ (with $i,j=L,R$) are known.
These are obtained by solving the Dyson equation in Nambu-harmonics space, which
relates the dressed Green functions to the bare ones and the self-energy $\hat{\Sigma}$. 
Details and solution of the Dyson equation are given in Appendix~\ref{DysonApp}. Note that the bare Green function of a superconductor in Nambu-harmonics space can readily be expressed in frequency representation~\cite{doniach1998a} as
\begin{equation}
	g^{r/a}_{n m} (\omega) = - \lim_{\delta\to 0}\frac{\omega_n \openone+\Delta\sigma_x}{\sqrt{\Delta^2-(\omega_n \pm i\delta)^2}} \delta_{nm} \, ,
\end{equation}
where $\omega_n = \omega + n \Omega$ and the large bandwidth limit is assumed without loss of generality.

Finally, it may turn out useful to compare the zero-frequency period averaged noise to a reference value. This is achieved by introducing a so-called excess noise (XN) which corresponds to the difference of the total noise, obtained when the junction is biased by a voltage with both a DC and AC components, to its purely DC counterpart
\begin{align}
S_\text{exc} = \left. \overline{\left\langle S_{LL}\right\rangle}\right|_{dc+ac}  - \left. \overline{\left\langle S_{LL}\right\rangle}\right|_{\text{DC}} \, .
\label{eq:ExcessNoise}
\end{align}

\section{Results in the limiting regimes} \label{sec:limits}

We now consider the results obtained by evaluating the current and the noise
using Eqs.~\eqref{Current_Fourier_global}, \eqref{Noise_Fourier_global} and \eqref{eq:ExcessNoise}, in the two limiting regimes of a very small and a very large gap compared to the driving frequency. Focusing on these limiting regimes allows us to apprehend more easily the physical processes at play in the junction, and even obtain analytic formulas in the extreme situations of a vanishing or an infinite superconducting gap. In the end, understanding these two limiting regimes helps us interpret the results obtained in the general case, described in the next section.

\subsection{Small-gap regime, $\Delta\ll\Omega$}\label{NormalSec}

This regime corresponds to the case of a driving frequency much larger than the gap. One thus expects that transport processes through the junction are largely dominated by QP-transfer. Indeed, the Andreev reflection of an incoming electron of energy $\omega$ (with respect to the chemical potential of the superconductor) occurs with a probability~\cite{blonder1982a,cuevas1999a} $\Delta^2/\omega^{2}$ and is thus strongly suppressed in this limit.

Let us first consider the extreme situation of a vanishingly small superconducting gap, which has the benefit of being tractable analytically. In this limit, the normal-metal and the BCS lead Green functions are identical and read
\begin{equation}
	\hat{g}_{\text{BCS}}^{r/a} \xrightarrow[\Delta \to 0]{} \hat{g}_{N}^{r/a} =\mp i \hat{\openone} \, .
\end{equation}
This, in turn, allows us to analytically solve the Dyson equation and thus derive the expression for both the current and the noise, yielding
\begin{equation}
	\overline{\left\langle I^N\right\rangle}_{q} = \frac{e}{\pi} \tau e V_{\text{DC}} = \frac{e\tau}{\pi} q\Omega\, ,
	\label{eq:Normal_currentIN}
\end{equation}
and
\begin{align}
		\overline{\left\langle S^N\right\rangle}_{q}= & \frac{e^2}{\pi} \bigg[4\tau^2\theta  +2\tau(1-\tau)\nonumber \\
		&  \times\sum_n (eV_{\text{DC}}+n\Omega)P_n(q)\coth \left(\frac{eV_{\text{DC}}+n\Omega}{2\theta} \right)\bigg] .
\label{eq:normal_noise}
\end{align}
Naturally, this limit corresponds to a simple junction between two normal-metals, hence the $N$ superscript in the above expressions. As expected, our microscopic model recovers the known results from scattering theory,~\cite{lesovik1994a} with the transmission coefficient $\tau$ naturally emerging from the microscopic tunneling constant $\lambda$ as $\tau=\frac{4\lambda^2}{(1+\lambda^2)^2}$.

The excess noise, expressed as a function of the charge $q$ injected per period by the voltage drive, then reads as
\begin{align}
		S_{\text{exc}}^N(q)  =  & \frac{e^2}{\pi}  2\tau(1-\tau) \Omega  \sum_n P_n (q) \nonumber \\
		&  \qquad \times \left\{  \frac{n+q}{\tanh \left[\frac{\left(n+q \right) \Omega}{2\theta} \right]} - \frac{q}{\tanh \left(\frac{q \Omega}{2\theta} \right)} \right\} 
\label{eq:normal_excessnoise}
\end{align}
A spectacular result is that for quantized Lorentzian pulses, most of the negative channels do not contribute at all to either the current or the noise, as $P_n(q)=0$ for any $n<-q$ (see Fig.~\ref{Floquet_Coefs} for an illustration and Ref. [\onlinecite{keeling2006a}] for a proof). It follows that, at low enough temperature, the PAN reduces to its DC value and the excess noise is fully suppressed for such quantized Lorentzian pulses.

\begin{figure}[t]
	\centering
	\includegraphics[width=0.48\textwidth]{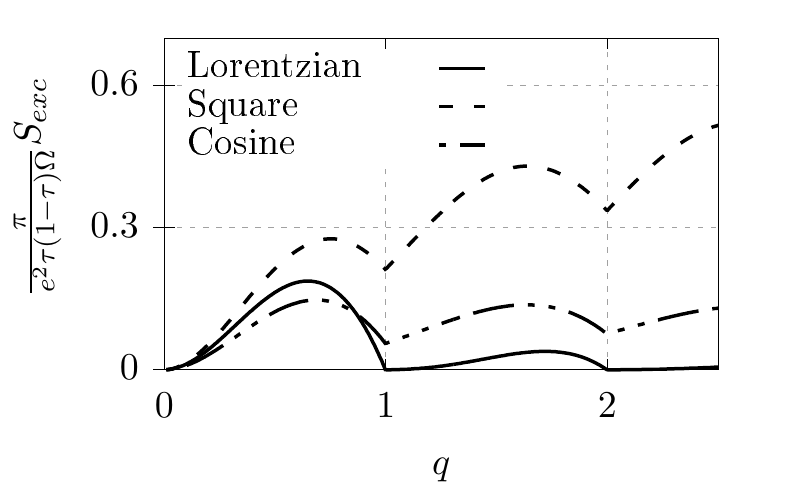}
	\caption{Excess noise for the $N-S$ junction driven by a cosine, a square and a periodic Lorentzian drive (with relative width $\eta=0.1$) as a function of the injected charge $q$ in the low gap regime, $\Omega=10^3\Delta$, at low temperature $\beta\Omega=10^3$. These plots are independent of the tunnel coupling but for the figure, we specified $\lambda=0.5$.}
	\label{Excess_noise_Normal}
\end{figure}

Going beyond this strict zero-gap limit, while still considering $\Delta \ll \Omega$, we could compute the excess noise numerically. Our results for the Lorentzian, cosine and square drives at low temperature are summarized in Fig.~\ref{Excess_noise_Normal}. For any drive, there is a local minimum of XN at integer $q=n$. This is most pronounced for a train of levitons for which the excess noise is still almost fully suppressed at integer $q$. Conversely, the sinusoidal bias features a residual finite XN for a quantized drive and the square drive bears local minima with even higher XN. These results are very much in agreement with previous works focusing on a normal junction~\cite{dubois2013a} signaling that the physics at play is well captured by considering only quasiparticle-transfer processes in this small-gap regime.

\subsection{Andreev regime, $\Delta \gg \Omega$}\label{AndreevSec}

We now turn to the opposite regime corresponding to either a large gap or a low driving frequency, i.e. $\Delta/\Omega \gg1$. There, we expect quasiparticle-transfer processes to become negligible as they demand a stiff price to pay in energy. Instead, Andreev reflection now becomes the leading mechanism involved in the transfer of charge through the junction, thus contributing to current and noise.

As a first step, let us consider the extreme situation of an infinite gap. This situation allows for a tractable analytical solution. Indeed, this amounts to replacing the BCS lead Green function by
\begin{equation}
	\hat{g}_{BCS}^{r/a} \xrightarrow[\Delta \to 0]{} \hat{g}_{A}^{r/a} =-\hat{\sigma}_x\, ,
\end{equation}
where the subscript $A$ stands for Andreev, as we are considering a limit where Andreev reflection is the only process available for scattering electron between the two leads.

After some algebra, we are able to solve analytically the Dyson equation and extract a full-fledged expression for the various dressed Green functions required to derive the current and the noise. Omitting the frequency dependence for convenience, these Green functions reduce to
\begin{widetext}
	\begin{align}
	\label{eq:Andreev_GLRpm}
			\hat{G}_{LR}^{\pm\mp} &=-\lambda\left\{\left[\hat{\mathcal{P}}^\dagger\hat{\sigma}_x \hat{\mathcal{P}},\hat{T}\right]\hat{\mathcal{P}}^\dagger\hat{\sigma}_x\frac{\tau_A}{4}\lambda^{-2} - i\frac{\tau_A}{4}\lambda^{-4}\hat{T}\hat{\mathcal{P}}^\dagger\hat{\sigma}_x \mp i\frac{\sqrt{\tau_A}}{2}\lambda^{-2} \hat{\mathcal{P}}^\dagger\hat{\sigma}_x-i\frac{\tau_A}{4} \hat{\mathcal{P}}^\dagger\hat{\sigma}_x\hat{\mathcal{P}}\hat{T}\hat{\mathcal{P}}^\dagger\right\}\, ,  \\
			\label{eq:Andreev_GLLpm}
			\hat{G}_{LL}^{\pm\mp} &=\left[\hat{\mathcal{P}}^\dagger\hat{\sigma}_x \hat{\mathcal{P}},\hat{T}\right]\frac{\sqrt{\tau_A}}{2}\left(1-\frac{\sqrt{\tau_A}}{2}\lambda^2\right) - i\left(1-\lambda^2\frac{\sqrt{\tau_A}}{2}\right)^2\hat{T} \mp i\left(\frac{\sqrt{\tau_A}}{2}\lambda^2-1\right)\hat{\openone} -i\frac{\tau_A}{4} \hat{\mathcal{P}}^\dagger\hat{\sigma}_x\hat{\mathcal{P}}\hat{T}\hat{\mathcal{P}}^\dagger\hat{\sigma}_x \hat{\mathcal{P}}\, , \\
			\label{eq:Andreev_GRRpm}
			\hat{G}_{RR}^{\pm\mp} &= -i\lambda^{-2}\frac{\tau_A}{4}\left\{ \hat{\sigma}_x\hat{\mathcal{P}}\hat{T}\hat{\mathcal{P}}^\dagger\hat{\sigma}_x - i\lambda^2\left[\hat{\sigma}_x,\hat{\mathcal{P}}\hat{T}\hat{\mathcal{P}}^\dagger\right] +\lambda^4\hat{\mathcal{P}}\hat{T}\hat{\mathcal{P}}^\dagger \mp \left(1+\lambda^4\right)\hat{\openone} \right\}  \, ,
	\end{align}
\end{widetext}
where $\hat{\mathcal{P}}$ is given by Eq.~\eqref{Pcaldef} and $\hat{T}$ has matrix elements of the form $T_{nm}=\delta_{nm}\tanh \left(\frac{\omega+n\Omega+\sigma_z eV_{\text{DC}}}{2\theta} \right)$.  An interesting quantity which naturally arises in this calculation is the so-called Andreev transmission $\tau_A$, which is related to the microscopic tunneling constant $\lambda$ as $\tau_A=\frac{4\lambda^4}{(1+\lambda^4)^2}$. This quantity mirrors the normal transmission introduced in the previous section upon replacing $\lambda$ with $\lambda^2$, as the basic scattering process is now Andreev reflection which involves the conversion of an incident electron into a backscattered hole, and therefore two tunneling events through the junction.

\begin{figure}[t]
	\includegraphics[width=0.48\textwidth]{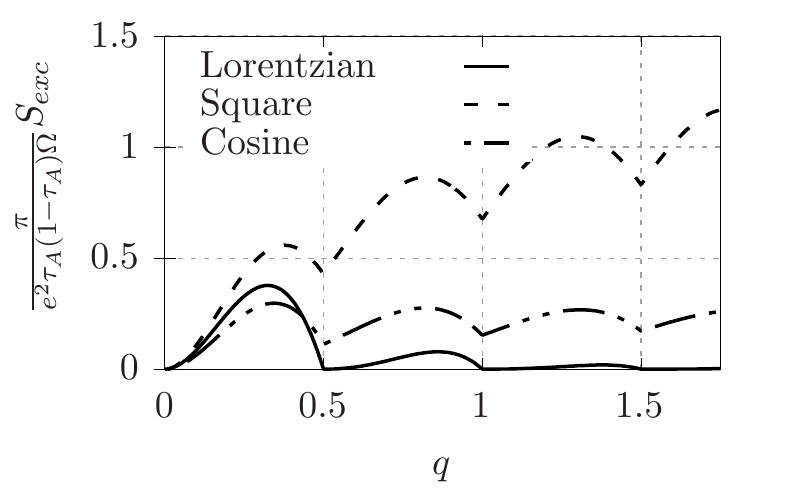}
	\caption{Excess noise for the $N-S$ junction driven by a cosine, a square and a periodic Lorentzian drive (with relative width $\eta=0.1$) as a function of the injected charge $q$ in the Andreev regime $\Omega=10^{-3}\Delta$, at low temperature $\beta\Omega=10^3$. These plots are independent of the tunnel coupling but for the figure, we specified $\lambda=0.5$.}
	\label{Excess_noise_Andreev}
\end{figure}

It is important to stress out that obtaining such self-contained expressions for the fully dressed Green functions implies that we can obtain analytical results for the current and noise to all orders in the tunneling Hamiltonian, at arbitrary temperatures, in this infinite gap limit.

Indeed, plugging Eq.~\eqref{eq:Andreev_GLRpm} back into Eq.~\eqref{Current_Fourier_global}, the average current can be readily computed leading to
\begin{equation}
	\overline{\left\langle I^A \right\rangle}_{q}=\frac{e\tau_A}{\pi}2q\Omega\, ,
	\label{eq:IAndreev}
\end{equation}
which displays an Ohmic behavior. Interestingly, as evidenced by Eq.~\eqref{eq:Normal_currentIN}, the above expression for the current in the Andreev limit corresponds exactly to the one of a $N-N$ junction with bias $2V (t)$ and a coupling constant $\lambda^2$, further illustrating the contribution from pure AR processes.

Similarly, substituting Eqs.~\eqref{eq:Andreev_GLRpm}-\eqref{eq:Andreev_GRRpm} into Eq.~\eqref{Noise_Fourier_global}, leads, after some cumbersome but straightforward manipulations, to the analytic expression for the PAN
\begin{align}
		\overline{\left\langle S^A \right\rangle}_{q} &= \frac{e^2}{\pi}\bigg[4\tau_A^2\theta +2\tau_A(1-\tau_A)                                                  \nonumber     \\
		&  \times\sum_n  (2eV_{\text{DC}}+n\Omega)P_n^A (q) \coth \left( \frac{2eV_{\text{DC}}+n\Omega}{2\theta} \right)\bigg]\, ,
	\label{Eq_Noise_andreev}
\end{align}
where an effective Floquet weight of the form
\begin{equation}
P_n^A(q)=\left\lvert\sum_s p_{n-s}(q) p_s(q) \right\rvert^2
\label{eq:FloquetAndreev}
\end{equation}
naturally appears in the calculation.
As it turns out, this set of Floquet weights actually correspond to a doubling of the applied voltage bias, as one can easily show that $P_n^A(q)=P_n(2q)$, reflecting once more the importance of Andreev reflection processes in this limit.

It follows that, just like the average current above, the PAN of the $N-S$ junction in the Andreev regime is related to that of a normal junction, yielding
\begin{equation}
	\overline{\left\langle I^A\right\rangle}_{q}=\overline{\left\langle I^N\right\rangle}_{q^*}
	\text{  and  }
	\overline{\left\langle S^A \right\rangle}_{q}=\overline{\left\langle S^N \right\rangle}_{q^*},
	\label{eq:equivNA}
\end{equation}
where $q^*=2q$ and all transmission coefficients $\tau$ in normal ${}^N$ expressions are now given by their Andreev counterpart $\tau_A$. This emphasizes that, in the Andreev regime, both averaged current and PAN can be interpreted as a joint tunneling of two electrons (with opposite spin), amounting to BCS pair creation in the superconducting junction through perfect AR. It follows that one expects non-trivial signatures to appear in the excess noise for half-integer values.

Building on this understanding, we now consider a large (but not infinite) gap, compared to the driving frequency. Results in this regime have to be obtained numerically, and are displayed in Fig.~\ref{Excess_noise_Andreev}  where we perform numerical computations of the excess noise for same three drives as before, the cosine, square and periodic Lorentzian drives.

As anticipated from the infinite gap limit, our results for the XN now display new local minima compared to the small-gap regime, located at half-integer values of the injected charge independently of the drive. For the square drive,  the XN globally increases with $q$, with sharply marked local minima at half-integer values of $q$. The cosine drive typically yields lower values of XN compared to the square drive, with softer local minima As before, the most interesting signature comes from the periodic Lorentzian drive, as for integer values of $2q$ the excess noise is almost fully suppressed, i.e., the PAN exactly reduces to its DC level counterpart for  Lorentzian pulses with half-integer injected charge. 

Compared with the low-gap regime, we analyze the present results as an effective doubling of the charge of the carriers exchanged at the junction, a direct consequence of the prominent role of Andreev reflection in this regime. This effective doubling appears at various stages throughout the calculation, and is most visible in the Floquet weights Eq.~\eqref{eq:FloquetAndreev}, as well as in the substitution $q^* = 2 q$ illustrated in Eq.~\eqref{eq:equivNA} when it comes to comparing the normal and the Andreev regimes.

The main feature of this large-gap regime is that a train of half quantized Lorentzian pulses leads to the injection of an integer number of Cooper pairs in the superconductor. These so-called ``Andreev levitons'' correspond to the minimal excitation states of the $N-S$ junction.

\section{Results in the general case}\label{InterSec}

We now consider the results in the regime where the drive frequency is comparable to the
superconducting gap $\Delta \sim \Omega$.
Our aim is to describe precisely the crossover behavior between the quasiparticle-transfer-dominated regime (small-gap case, $\Delta \ll \Omega$) and the Andreev regime (large-gap case, $\Delta \gg \Omega$). In this situation, analytical calculations are for the most part untractable as it becomes impossible to solve the Dyson equation exactly, forcing us to resort either to expanding analytic expressions perturbatively in $\lambda$ or to performing exact numerical calculations, i.e., including contributions at all orders in the tunneling constant. For clarity, some intermediate steps of the calculations are presented in Appendix~\ref{CalculationApp}.

\subsection{Effective gaps}

The analysis we carried out in the previous section showed that in the small-gap regime, the transport properties are dominated by quasiparticle transfer outside the gap, while in the large-gap regime, Andreev reflection processes, which occur inside the gap, are the main driving mechanism. In the intermediate regime, where $\Omega \sim \Delta$, we thus expect transport to be impacted by both quasiparticle transfers outside the gap and Andreev reflection processes inside the gap, as well as interference between the two. However, because of the Floquet decomposition, every channel is confronted with a different gap in energy when compared with its own effective chemical potential. Indeed, as each Floquet channel (with label $n$) corresponds to a Fermi sea shifted by $eV_{\text{DC}} + n \Omega$, the gap for the channel $n$ is spanned by a DC voltage in the range $[-\Delta - n \Omega, +\Delta - n\Omega]$. Quite importantly, this means that for a given applied DC voltage, some channels are mainly in the Andreev regime, while others are in the quasiparticle-dominated regime. The average current and noise then result from the scattering between these channels.

To better understand the features we observe, it is convenient to introduce an effective gap associated with a given Floquet channel $n$ as
\begin{equation}
	[\gamma_n^{-},\gamma_n^{+}] \mbox{ ,     with      } \gamma_n^\pm=-n\pm \frac{\Delta}{\Omega}
\end{equation}
Since we are mostly interested in the evolution of the transport properties as a function of $q$, which represents the DC bias in units of the driving frequency, the above effective gap is also expressed as a range in energy, given in units of $\Omega$. Within this frame, it corresponds to the range of the injected charge per period $q$ for which the Floquet channel $n$ sees the superconducting gap $\Delta$. 

Note that the width of the effective gap is $2\Delta/\Omega$,
which naturally implies that the effective gaps of the different channels are large and overlapping
when $\Omega < \Delta$, while they are small and well separated for $\Omega > \Delta$.

\subsection{Average current}

A closed-form analytical expression for the average current is not available in the general case. However, one can still work out an intermediate form which turns out to be quite useful. Indeed, after some lengthy derivation (additional details are provided in App.~\ref{CalculationApp}), one has for the average current
\begin{align}
		\overline{\left\langle I \right\rangle}_{q} &= 2e\lambda^2\sum_{n} P_{n}(q) \int_{-\infty}^{\infty}\frac{\mathrm{d}\omega}{2\pi} \, {\cal I}(\omega) \nonumber \\
& \times \left[\tanh\left(\frac{\omega - (n+q) \Omega}{2\theta}\right) - \tanh\left(\frac{\omega + (n+q) \Omega}{2\theta}\right)\right],
\label{eq:avgI}
	\end{align}
where we introduced
\begin{equation}
{\cal I} (\omega)    = \left\{
\begin{aligned}
			\frac{2\lambda^2}{\left(1+\lambda^4 \right)^2 \left(1 - \frac{\omega^2}{\Delta^2} \right) + 4 \lambda^4 \frac{\omega^2}{\Delta^2}}                 & \qquad\text{if}\quad\left\lvert\omega\right\rvert<\Delta     \\
			\frac{1}{\left( 1 + \lambda^4 \right) \sqrt{1 - \frac{\Delta^2}{\omega^2}}+2 \lambda^2} & \qquad\text{if}\quad\left\lvert\omega\right\rvert>\Delta\, .
\end{aligned}\right.
\label{eq:Icalomega}
\end{equation}
In the presence of an AC drive, the average current Eq.~\eqref{eq:avgI} is therefore the sum of independent contributions coming from all Floquet channels. 

\begin{figure}[t]
	\centering
	\includegraphics[width=0.48\textwidth]{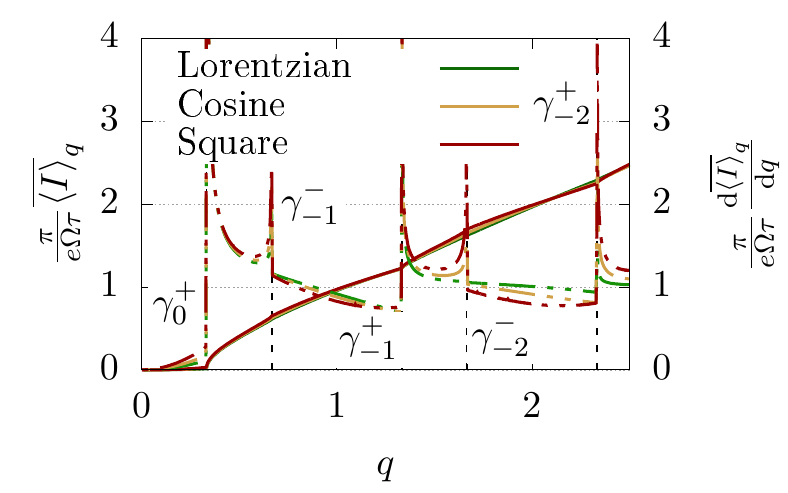}
	\caption{The average current (full line) and its derivative with respect to $q$ (dashed line) as the junction is driven by different drives (Lorentzian with $\eta=0.15$, cosine voltage and square voltage), at $\Omega=3\Delta$, $\beta\Omega=10^3$ and for a tunnel barrier $\lambda=10^{-2}$, as a function of $q$. The edges of the effective gap for conducting channels $n=0,1$ and $2$ are shown by vertical dashed lines.}
	\label{Average_Current_Inter}
\end{figure}

Interestingly, the behavior of the resulting integral is strongly impacted by the behavior of $\left| \omega \pm (n+q) \Omega\right|$ with respect to the superconducting gap $\Delta$, signaling the importance of the effective gap  $[\gamma_n^-;\gamma_n^+]$ associated with each Floquet channel $n$. In particular, the integrand of Eq.~\eqref{eq:avgI} changes sharply for values of the DC voltage near $\gamma_n^\pm$, so that one would expect the average current to undergo a sudden change of behavior close to these values. As a consequence, the transition from a purely AR-dominated behavior to a QP-transfer-dominated current should then be visible for each channel in the total average current.

This is illustrated in Fig.~\ref{Average_Current_Inter} where we plot both the average current and the differential conductance as a function of the injected charge $q$, for $\Omega = 3 \Delta$ and different drives. To insist some more on this transition, we chose here to focus on a low-transparency tunnel barrier, where Andreev processes are strongly suppressed as they involve two tunneling events (thus higher order in $\lambda$). 

The average current looks smooth exhibiting only small kinks near the effective gaps edges and an overall behavior which looks rather independent of the details of the drive. At low voltage, the current is vanishingly small then suddenly increases near $q \simeq \gamma_0^+$, as one goes from an almost no current regime (AR-dominated) to one that is dominated by QP-transfer. These signatures are much more striking in the differential conductance as they manifest as sharp peaks located at the effective gap edges of the different Floquet channels, which are only softened by the finite temperature. Within a given effective gap or in-between gaps, the differential conductance remains continuous and featureless.

\subsection{Noise}

The general analytical calculation of PAN turns out to be quite cumbersome and is not shown here. Instead, we rely on two rather complementary approaches. First, we focus on a perturbative expansion in the tunneling constant. Indeed, an expansion in powers of $\lambda$ (with only even powers contributing) can be performed and the first few terms are accessible analytically. The general expression reads as 
\begin{equation}
	\overline{\left\langle S \right\rangle}_{q}=\sum_{n=1}^\infty\overline{\left\langle S \right\rangle}_{q}^{(2n)}\, ,
\end{equation}
where $\overline{\left\langle S \right\rangle}_{q}^{(2n)}$ is the term of order $O \left( \lambda^{2n} \right)$. Here we compute only the $n=1$ term exactly and the contribution due to in-gap AR of the $n=2$ term.
This helps better understand the full picture, as in the low-transparency regime, this allows one to isolate the leading contribution from QP-transfer $\left[ \text{of order }O \left( \lambda^2 \right) \right]$ and the main modifications arising from Andreev reflection $\left[ \text{only present from order } O \left( \lambda^4 \right) \text{ onward} \right]$. 
The second approach consists in solving numerically the Dyson equation in order to obtain the evolution of the excess noise as a function of the applied drive. This provides a complete panorama of the different competing physical processes at play in the junction.

\subsubsection{Order $\lambda^2$ and negative excess noise}\label{Neglambda2}
Since the expression for the noise already contains a prefactor of order $\lambda^2$, the computation of $\overline{\left\langle S \right\rangle}_{q}^{(2)}$ is carried out by replacing all the dressed Green functions in Eq.~\eqref{Noise_Fourier_global} by their bare equivalent. The second order component of the noise can then be immediately written as
\begin{align}
	\overline{\left\langle S \right\rangle}_{q}^{(2)}=& 2 e^2 \lambda^2 \sum_n P_{n}  \int_{|\omega| > \Delta} \frac{d\omega}{2\pi} 
	\frac{|\omega|}{\sqrt{\omega^2 - \Delta^2}} \nonumber \\
	&\times \left[ 1 - \tanh \left( \frac{\omega + eV_{\text{DC}} + n\Omega}{2 \theta} \right) \tanh \left( \frac{\omega}{2 \theta} \right)  \right]  \, .
	\label{lambda2noise}
\end{align}
It naturally arises from the calculation that this contribution only describes QP-transfer above the gap as all the non-diagonal terms in Nambu space from the bare superconducting lead Green function do not contribute in the end. As it turns out, the resulting expression bears some striking resemblance with the corresponding $\lambda^2$ contribution from the DC noise $S_{\text{DC}}^{(2)} (eV)$, allowing us to write
\begin{equation}\label{lambda2noiseTG}
	\overline{\left\langle S \right\rangle}_{q}^{(2)}=\sum_n P_{n} S_{\text{DC}}^{(2)} \left( e V_{\text{DC}} + n \Omega \right) \, .
\end{equation}
This intriguing feature is deeply connected with the works of Tien and Gordon~\cite{tien1963a}. We investigate it in more details in the next section.

\begin{figure}[t]
	\centering
	\includegraphics[width=0.48\textwidth]{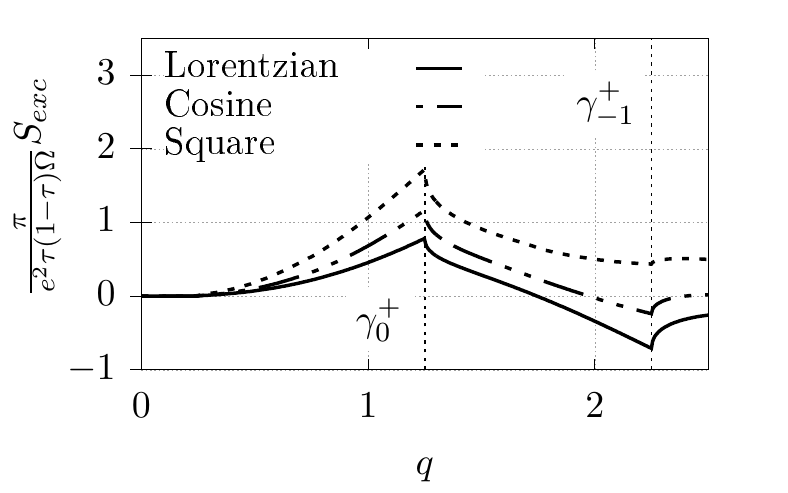}
	\caption{The XN as a function of the injected charge $q$ for three different drives (Lorentzian with $\eta=0.15$, cosine voltage and square voltage), at frequency $\Omega=0.8\Delta$, $\beta\Omega=10^3$ and at low transparency ($\lambda=10^{-2}$).}
	\label{Neg_Excess_noise}
\end{figure}

This connection with DC noise allows us to better understand one of the surprising results obtained at low transparency for intermediate frequency. This is illustrated in Fig.~\ref{Neg_Excess_noise}, where one readily sees that in such a regime the excess noise can become negative, thus signaling that adding an AC component to a DC drive can lead to a reduction of the noise for an appropriately chosen set of parameters. Indeed, while the cusps near $q = \gamma_n^+$ (visible in Fig.~\ref{Neg_Excess_noise} for all types of drives) are reminiscent of what was observed for the current, and can be understood in terms of a transition between AR and QP dominated regimes, the presence of negative excess noise for the Lorentzian and the cosine drives is more intriguing (we could not find a regime where this was also true for the square drive). 

This can be understood with the help of Eqs.~\eqref{lambda2noise} and \eqref{lambda2noiseTG} where one notices that if the contribution of the different Floquet channels is non-linear in voltage, the PAN is not necessarily equal to or larger than its DC counterpart. In practice, this non-linearity occurs primarily near the edges of the gaps $[\gamma_n^-;\gamma_n^+]$, which precisely corresponds to the regions of negative XN. Focusing on just a subset of channels, considering only $n=-1$, $0$ and $1$, and noticing that the DC noise is a monotonically increasing concave function of the voltage (for $eV_{\text{DC}}>\Delta$) it becomes quite easy to find a set of $P_n$ satisfying both $\sum_n P_n = 1$ and
\begin{align}
 P_{-1}  S_{\text{DC}} \left( eV_{\text{DC}}-\Omega \right) &+ P_{1} S_{\text{DC}} \left( eV_{\text{DC}}+\Omega \right) \nonumber \\
& \qquad < \left( 1 - P_0 \right) S_{\text{DC}} \left( eV_{\text{DC}} \right)\, ,
\end{align}
which, in turn, leads to a negative value of the excess noise at this particular voltage. From this, it follows that the negative excess noise arises from the convexity properties of the DC noise, which is itself a consequence of the gap edge singularity in the spectrum of quasiparticles from the superconducting lead.

Finally, we feel important to stress out that this reduction of the PAN below the DC level should be experimentally detectable for cosine and Lorentzian drives.

\subsubsection{Excess Andreev noise to order $\lambda^4$}

For voltages within the gap of a given Floquet channel, the PAN is only due to correlations between Andreev and normal reflections. The lowest order for such correlations is $\lambda^4$ and can be obtained analytically after some lengthy but straightforward algebra (additional details are presented in App.~\ref{DysonApp} and \ref{CalculationApp}). Indeed, for energies lying inside the effective gap of a particular channel, the retarded and advanced bare Green functions of the superconducting electrode are equal so that the corresponding greater ($g^{+-}$) and lesser ($g^{-+}$) Green functions vanish. However, since normal reflection also plays an important role in this regime, the calculation no longer simplifies into an expression involving effective Floquet weights, as it did in the Andreev limit of infinite gap, [see Eq.~\eqref{eq:FloquetAndreev}]. Instead, in this situation, the $\lambda^4$ PAN contribution involving AR is a sum over three harmonics indices and reads as
\begin{widetext}
	\begin{align}
			\overline{\left\langle S\right\rangle}_{\text{Andreev}}^{(4)} & =32\int_{-\infty}^{\infty}\frac{\mathrm{d}\omega}{2\pi}\sum_{nsr=-\infty}^{+\infty}\lambda^2\Delta_{n}\lambda^2\Delta_{r}p_{n}^*p_{r}p_{s-n}^*p_{s-r}
			\Theta\left(\Delta-\left\lvert\omega+n\Omega\right\rvert\right)
			\Theta\left(\Delta-\left\lvert\omega+r\Omega\right\rvert\right)\times                                                                                                                                                                                                                                \nonumber \\
			                                                               & \qquad\times\left\{f\left(\frac{\omega -(e V_{\text{DC}} + s\Omega)}{2}\right)\left[1-f\left(\frac{\omega-e V_{\text{DC}} }{2}\right)\right]+f\left(\frac{\omega-e V_{\text{DC}} }{2}\right)\left[1-f\left(\frac{\omega -(e V_{\text{DC}}  + s\Omega)}{2}\right)\right]\right\}\, ,
			                                                               \label{lambda4noise}
	\end{align}
\end{widetext}
where $\lambda^2\Delta_n=\lambda^2 \frac{\Delta}{\sqrt{\Delta^2-(\omega+n\Omega)^2}}$ is the AR amplitude of the Floquet channel $n$, and $\Theta(x)$ is the Heaviside distribution.

This term cannot be further reduced into a simpler form only involving Floquet weights, as obtained in Eq.~\eqref{lambda2noiseTG} for the $O \left( \lambda^2 \right)$ contribution to the noise. This means that, in this intermediate regime, the noise can no longer be interpreted as a sum of independent contributions coming from all Floquet channels. Instead, the PAN now involves products of two types of terms, for example $\lambda^2\Delta_r p_r p_{s-r}$, which describes the interference between two transport events, Andreev and normal reflections.

\begin{figure*}
	\centering
	\includegraphics[width=0.965\textwidth]{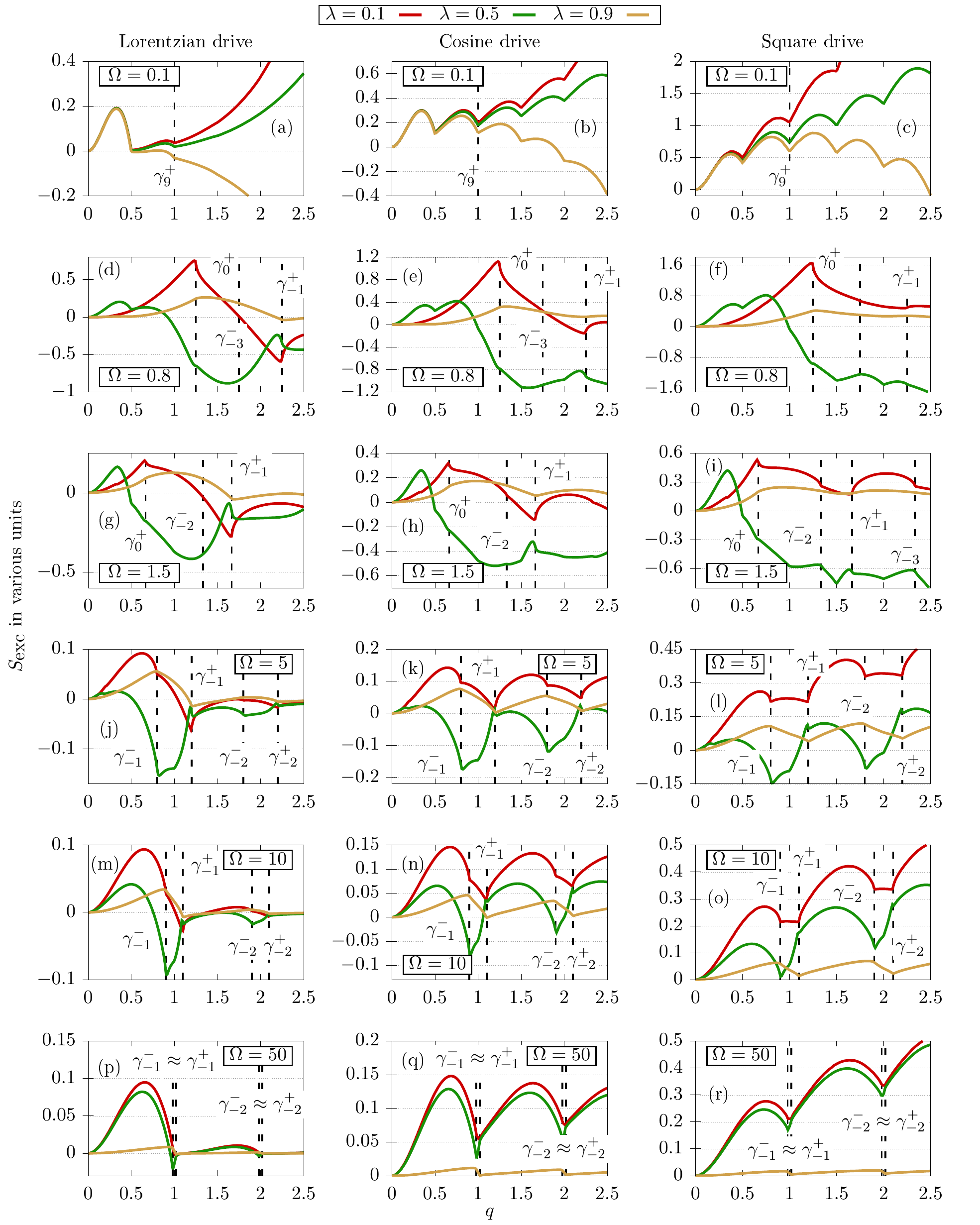}
	\caption{The excess noise  at an $N-S$ junction for three different drives and different transparencies as a function of the injected charge $q$ for low temperature: $\beta\Omega=10^3$. Warning: the curves are not all normalized in the same way. For $\Omega=0.1\Delta$, i.e. plots (a)-(c), they are all in units of $e^2\Omega\tau_A(1-\tau_A)/\pi$, for $\Omega\neq0.1 \Delta$, the XN is in units of $e^2\Omega\tau(1-\tau)/\pi$ \textit{except for the case $\lambda=0.9$ where the XN is in units of $e^2\Omega\tau_A\tau/\pi$.}}
	\label{Excess_Noise_Inter_Tot}
\end{figure*}

\subsubsection{Excess noise for all transparencies}

We now turn to a detailed quantitative investigation of the PAN, whose behavior can be probed through an XN analysis. Using Eqs.~\eqref{Noise_Fourier_global} and \eqref{eq:ExcessNoise}, we numerically obtain the excess noise in a wide range of parameters, from low to high transparency, from frequencies smaller to larger than the gap and for three different drives (cosine, square, and Lorentzian). The corresponding plots are displayed in Fig.~\ref{Excess_Noise_Inter_Tot}.

As explained in the caption of Fig.~\ref{Excess_Noise_Inter_Tot}, in order to display the curves on the same graph, we had to adjust normalizations for the XN as follows. For $\Omega=0.1\Delta$ [Figs.~\ref{Excess_Noise_Inter_Tot}(a)--\ref{Excess_Noise_Inter_Tot}(c)] the normalization is $\tau_A(1-\tau_A)\Omega/\pi$ because this constitutes the Andreev-dominated regime. For $\Omega\neq0.1\Delta$ [Figs.~\ref{Excess_Noise_Inter_Tot}(d)--\ref{Excess_Noise_Inter_Tot}(r)] the XN is normalized by $\tau(1-\tau)\Omega/\pi$ for low ($\lambda=0.1$) and intermediate ($\lambda=0.5$) transparencies, because we aim at characterizing the transition to the normal-metal regime. Finally, for $\Omega\neq0.1\Delta$ and high transparency ($\lambda=0.9$), i.e., the red curves in Figs.~\ref{Excess_Noise_Inter_Tot}(d)--\ref{Excess_Noise_Inter_Tot}(r), the XN is normalized by $\Omega\tau_A\tau/\pi$ as it now involves both AR and QP-transfer.

Let us start with some general observations. For drive frequencies much smaller than the superconducting gap $\Delta$ [Figs.~\ref{Excess_Noise_Inter_Tot}(a)--\ref{Excess_Noise_Inter_Tot}(c)], all signals, at all transparencies, exhibit a minimum at $q=1/2$, depicted by ``arches'' with a positive XN. For the Lorentzian drive, this first arch is characterized by a fully suppressed excess noise at $q=1/2$, which constitutes the half-leviton regime. These minima typically persist for larger $q=n/2$ ($n$ integer), although they get less and less visible as the amplitude of the arches quickly decreases. In the opposite limit of frequencies much larger than the gap [Figs.~\ref{Excess_Noise_Inter_Tot}(p)--\ref{Excess_Noise_Inter_Tot}(r)], the minima still occur for all signals at all transparencies, only they are now located at integer $q$, mimicking the XN of a normal-metal junction. These integer minima yield a zero XN only for the Lorentzian drive.

\paragraph{Tunneling regime.}

Let us first focus on the tunneling regime, corresponding to all red curves in Fig.~\ref{Excess_Noise_Inter_Tot} (associated with $\lambda=0.1$). 

In the low-frequency regime [Figs.~\ref{Excess_Noise_Inter_Tot}(a)--\ref{Excess_Noise_Inter_Tot}(c)] , the XN is characterized by arches with local minima for half-integer values of $q$, on top of a monotonously increasing background contribution. For the Lorentzian drive, and unlike the other two, the amplitude of these arches quickly vanishes making them barely visible beyond $q=1$, while they look much more robust for cosine and square drives. As argued in the previous section on the Andreev regime (see Sec.~\ref{AndreevSec}), these signatures are typically associated with the preeminent role of AR processes.

Increasing the driving frequency to values comparable with the superconducting gap  [$\Omega=0.8\Delta$ in Figs.~\ref{Excess_Noise_Inter_Tot}(d)--\ref{Excess_Noise_Inter_Tot}(f), $\Omega=1.5\Delta$ in Figs.~\ref{Excess_Noise_Inter_Tot}(g)--\ref{Excess_Noise_Inter_Tot}(i)], these half-integer arches get completely washed out and the only remaining features in the XN are cusps located at the edges of the Floquet channel effective gaps. This change of behavior is attributed to the onset of the contribution of QP-transfer. Indeed, for such intermediate frequencies, QP-transfer becomes important, ultimately drowning out the contribution from AR processes which involve higher order terms in the tunneling constant. Negative XN might be visible for the Lorentzian and cosine drives, as already explained in Sec.~\ref{Neglambda2}.

Increasing further the driving frequency [$\Omega=5\Delta$ in Figs.~\ref{Excess_Noise_Inter_Tot}(j)--\ref{Excess_Noise_Inter_Tot}(l), $\Omega=0\Delta$ in Figs.~\ref{Excess_Noise_Inter_Tot}(m)--\ref{Excess_Noise_Inter_Tot}(o)] leads to the resurgence of arches, only now extending beyond half-integer values of $q$ and all the way to the edge of the effective gaps. For very large frequencies [Figs.~\ref{Excess_Noise_Inter_Tot}(p)--\ref{Excess_Noise_Inter_Tot}(r)], these gap edges tend to merge at integer values of the injected charge. The resulting XN then fully corresponds to that of a normal junction, recovering the results of Sec.~\ref{NormalSec} and identically matching the predictions of Ref. [\onlinecite{dubois2013b}].

It is important to stress out that these integer arches observed at $\Omega \gg \Delta$ are not a deformed version of the half-integer ones obtained for $\Omega \ll \Delta$, as the two sets rely on very different physical mechanisms, namely QP-transfer and Andreev reflection, respectively. Tuning the driving frequency favors one process over the other, leading to the corresponding set of local minima.

\paragraph{Intermediate transparency regime.}

This regime corresponds to the green curves in Fig.~\ref{Excess_Noise_Inter_Tot} (associated with $\lambda=0.5$). 

As in the low-transparency regime, we observe arches with local minima for half-integer $q$ at low frequency [$\Omega = 0.1\Delta$, Figs.~\ref{Excess_Noise_Inter_Tot}(a)--\ref{Excess_Noise_Inter_Tot}(c)]. These disappear at higher frequency, giving way to marked structures at the edges of the effective gaps of the Floquet channels, which then evolve into another set of arches extending between $\gamma_{n-1}^+$ and $\gamma_{n}^-$. At very high driving frequency [Figs.~\ref{Excess_Noise_Inter_Tot}(p)--\ref{Excess_Noise_Inter_Tot}(r)], these gap edges satisfy $\gamma_n^-\approx\gamma_n^+$ and $\gamma_{n+1}^{\pm}-\gamma_{n}^\pm\approx1$ and we recover for all drives, the arches with minima at integer $q$ as expected from our results in Sec.~\ref{NormalSec}.

\begin{figure}
	\centering
	\includegraphics[width=0.48\textwidth]{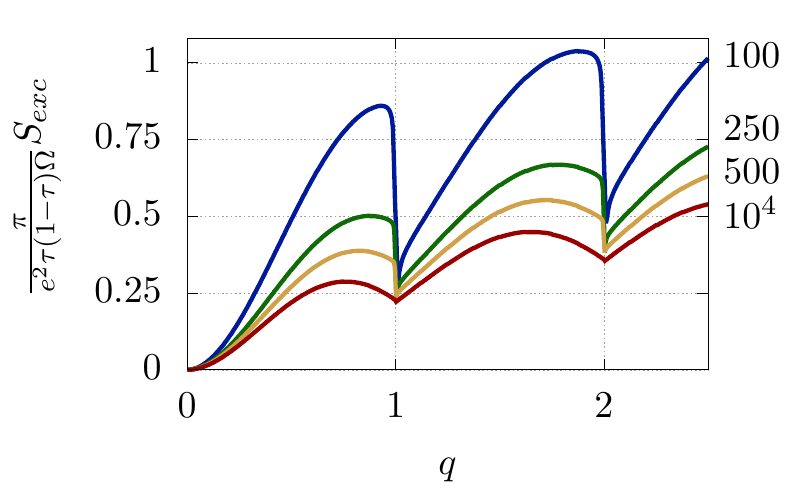}
	\caption{The Excess noise at the highly transparent junction ($\lambda=0.9$) driven by a square drive of different frequencies (specified on the right of the curves), as a function of the injected charge and at low temperature: $\beta\Omega=10^3$.}
	\label{Excess_Noise_Trans}
\end{figure}

However, this regime also differs from the low-transparency case in two important ways. First of all, one needs to go to much higher frequency in order to see the half-integer minima completely disappear, as remnants of these arches can still be observed at frequencies as high as $1.5\Delta$ [Figs.~\ref{Excess_Noise_Inter_Tot}(g)--\ref{Excess_Noise_Inter_Tot}(i)]. We interpret this as being related to the higher value of the tunneling constant, which thus allows Andreev reflection processes to remain quite important even at relatively high frequency. Secondly, it appears that negative excess noise can be observed over much larger regions of the bias voltage, and for all considered drives [Figs.~\ref{Excess_Noise_Inter_Tot}(d)--\ref{Excess_Noise_Inter_Tot}(n)]. Again, we believe this has to do with the larger tunneling constant, as we argued previously that this reduction of the period averaged noise compared to its DC level counterpart was to be attributed to interference effects between AR and NR across different Floquet channels, which become much stronger at intermediate transparency.

\paragraph{Quasi-transparent junction.}

This corresponds to the set of orange curves in Fig.~\ref{Excess_Noise_Inter_Tot} (associated with $\lambda=0.9$). 

As in the previous two regimes analyzed, and in accordance with our results from Secs.~\ref{NormalSec} and \ref{AndreevSec}, the excess noise exhibits two different types of arch structures: one present at very low frequency [Figs.~\ref{Excess_Noise_Inter_Tot}(a)--\ref{Excess_Noise_Inter_Tot}(c)] with local minima located at half-integer $q$, and one at high frequency [Figs.~\ref{Excess_Noise_Inter_Tot}(p)--\ref{Excess_Noise_Inter_Tot}(r)] with minima located at integer values of the injected charge.

In-between these limiting regimes, however, the excess noise is rather featureless, only showing slight kinks near the edges of the effective gaps $\gamma_n^\pm$.
As it turns out, all the physics studied thus far, with the interplay between Andreev reflection and QP-transfer, is strongly renormalized by NR up to very high order, which tends to smooth all previously observed signatures.

The only notable feature of this regime lies in the very high frequency regime where, while arches clearly appear with local minima located at integer values of the injected charge $q$, they typically seem to be much more asymmetric than at lower transparency. This is made clearer in Fig.~\ref{Excess_Noise_Trans} where the XN is displayed for increasingly higher frequency, from $\Omega = 10^2 \Delta$ to $\Omega = 10^4 \Delta$. Here we focused solely on the square drive for illustrative purposes. From this, it seems that this asymmetry is somewhat related to Andreev reflection processes as the steepest parts of the plots correspond to the effective gaps $\left[ \gamma_n^- ; \gamma_n^+ \right]$, which ultimately shrink down to zero as the frequency increases further.

In the end, this may seem like the least interesting regime when looking at the characteristic signatures of the excess noise. However, as we argue in Sec.~\ref{OnDemand}, it may also be the most relevant one when it comes to practical applications.

\section{Tien-Gordon-type relations} \label{sec:TGlike}

As we already hinted in the previous two sections, there exists a strong connection between the transport properties of the $N-S$ junction biased simultaneously by DC and AC voltages, and that of the same device in the absence of AC modulation. Such a connection was first unveiled~\cite{tien1963a} by Tien and Gordon when studying the tunneling current in superconducting diodes biased by a sinusoidal drive of frequency $\Omega$. In this context, they could show that
\begin{align}
I_{ac+dc} \left( V_{\text{DC}} \right) = \sum_l J_l^2 \left(\frac{e V_1}{\Omega} \right) I_{\text{DC}} \left( V_{\text{DC}} + l \frac{\Omega}{e}\right)
\end{align}
so that the average current in the presence of a drive $V (t) = V_{\text{DC}} + V_1 \sin (\Omega t)$ is a weighted sum of voltage-shifted DC currents (with a shift quantized in the driving frequency), with coefficients which are directly related to the AC power.
This has since generated a substantial amount of related work where it was realized that such Tien-Gordon-type relations are valid for different transport properties in a variety of two-terminal nanostructures. Here we analyze to what extent such relations hold in the biased $N-S$ junction under investigation.

\subsection{Average current}

Let us start back from our derivation of the average current in the general case. Looking back at Eqs.~\eqref{eq:avgI} and \eqref{eq:Icalomega} and considering the strict DC regime (which amounts to replacing $P_n$ with $\delta_{n 0}$), one readily sees that the average current obeys a Tien-Gordon law of the form
\begin{equation}\label{TGInterI}
	\overline{\left\langle I \right\rangle}_{q}= \sum_n P_n (q)  I_{\text{DC}} \left( eV_{\text{DC}} + n \Omega \right) \, ,
\end{equation}
where the DC current agrees with the known results from the literature,~\cite{cuevas1999a}
\begin{align}
		 I_{\text{DC}} \left( eV_{\text{DC}} \right) & = 2 e \lambda^2 \int \frac{d\omega}{2 \pi} {\cal I}(\omega) \nonumber \\
& \times \left[\tanh\left(\frac{\omega - eV_{\text{DC}}}{2\theta}\right) - \tanh\left(\frac{\omega + eV_{\text{DC}}}{2\theta}\right)\right]  ,
\end{align}
and ${\cal I}(\omega)$ is given in Eq.~\eqref{eq:Icalomega}.

\begin{figure}[t]
	\centering
	\includegraphics[width=0.48\textwidth]{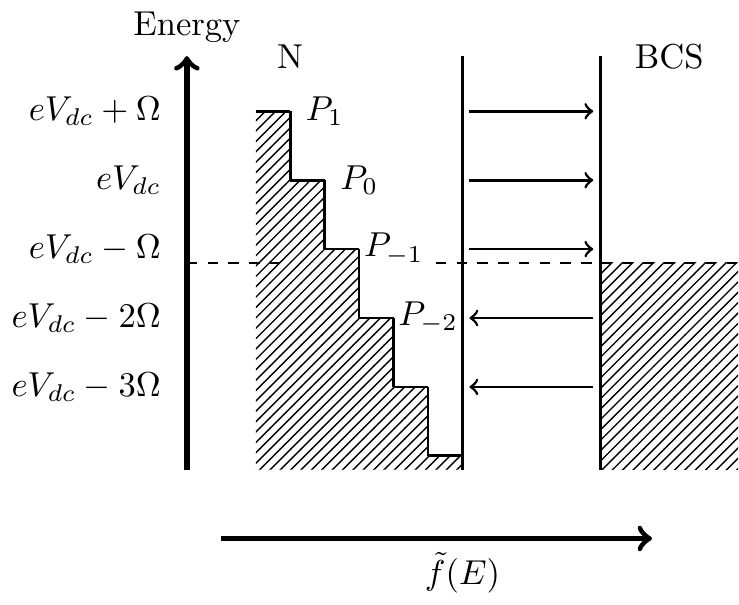}
	\caption{Schematic representation of tunneling events at the junction. Each channel, involving a Fermi sea at level $eV_{\text{DC}}+n\Omega$, contributes with a weight $P_n$ in the current, and is represented by a step in the figure. The arrows represent the direction of the electron current. The horizontal axis represents $\tilde{f}(E)$ which is defined in Eq.~\eqref{eq:tot_fermi_see}.}
	\label{Normal_limit}
\end{figure}

This result can actually be readily understood within Floquet theory. Indeed, as argued in Sec.~\ref{sec:floquet}, the voltage biased normal lead can be described as a set of Floquet channels, i.e., a superposition of Fermi seas shifted in energy with a weight given by the probability of absorbing or emitting a certain number of photons. From this, it ensues that the fraction of occupied states at a given energy in the metal is no longer given by the standard Fermi distribution but instead by the following weighted sum:
\begin{equation}
	\label{eq:tot_fermi_see}
	\tilde{f}(E)=\sum_{l=-\infty}^{\infty} P_l f(E-eV_{\text{DC}}-l\Omega)\, ,
\end{equation}
corresponding to the situation depicted in Fig.~\ref{Normal_limit}. Invoking scattering theory, one can then argue that the current only involves a linear combination of the leads distribution function, so that in the presence of an AC drive, the average current reduces to the sum of the DC contributions arising from all Floquet channels $n$, with the corresponding probability $P_n$. 

It is important to stress that this Tien-Gordon relation for the average current, Eq.~\eqref{TGInterI}, is a very general result, valid here for all regimes and any choice of the parameters of the $N-S$ junction (transmission, gap, temperature, frequency, voltage drive, etc.). It is trivially satisfied in the limiting regimes ($\Delta \to 0$ and $\Delta \to \infty$) where the current in the DC limit is linear in the applied voltage, as can be readily seen from Eqs.~\eqref{eq:Normal_currentIN} and \eqref{eq:IAndreev}.

\subsection{Period-averaged noise}

More interestingly, some of our results suggest the possibility to also establish a Tien-Gordon-type relation for the period-averaged noise.

\subsubsection{Limiting regimes}

In the limit of a vanishingly small gap $\Delta \to 0$, we are left with a junction between two normal-metals. Our expression for the noise, Eq.~\eqref{eq:normal_noise}, can be shown to satisfy
\begin{equation}\label{S_Normal_TG}
	\overline{\left\langle S^N \right\rangle}_{q}= \sum_n P_n  S^N_{\text{DC}} \left(eV_{\text{DC}}+n\Omega \right) \, ,
\end{equation}
where $S^N_{\text{DC}} \left(V \right)$ is the standard expression for the noise of a DC biased normal junction at finite temperature, and can be readily obtained from Eq.~\eqref{eq:normal_noise} by replacing $P_n$ with $\delta_{n0}$. In this normal regime, the PAN can thus be viewed as the sum of the contributions from independent Floquet channels separated in energy, in a similar way to the average current.

In the opposite limit of an infinite gap, it is also possible to relate the period-averaged noise to its DC driven counterpart, only it now involves a slightly modified form of the Tien-Gordon formula, namely,
\begin{equation}\label{S_Andreev_TG}
	\overline{\left\langle S^A \right\rangle}_{q} = \sum_n P_n^A
	  S^A_{\text{DC}} \left( eV_{\text{DC}}+n\frac{\Omega}{2} \right),
\end{equation}
where  $S^A_{\text{DC}} \left( eV \right)$ is the noise of the DC biased junction in the Andreev regime, which can be readily obtained from Eq.~\eqref{Eq_Noise_andreev} by replacing $P_n^A$ with $\delta_{n0}$. Note that this expression generalizes previously established zero-temperature results~\cite{belzig2016a} to finite temperature.

\begin{figure}[t]
	\includegraphics[width=0.48\textwidth]{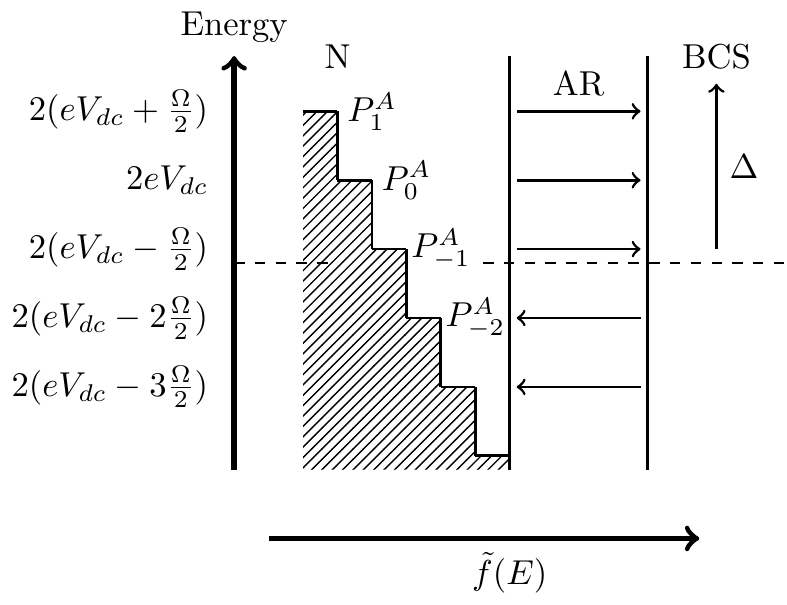}
	\caption{Symbolic representation of the junction as seen through tunneling current. The left metal is in a superposition of states which are superpositions of Fermi seas as described in the text and each step corresponds to an effective channel. The arrows indicate the direction of the electronic AR current.}
	\label{Andreev_Channels}
\end{figure}

This expression can be interpreted in terms of effective channels with weight $P_n^A$. These obviously do not correspond to the Floquet channels involved in the normal junction: the shift in energy corresponds to half the driving frequency and the Floquet weight $P_n^A$ consists of a superposition of the usual Floquet amplitudes $p_l(q)$. These effective channels are illustrated in Fig.~\ref{Andreev_Channels}.

Note that this representation, while very different from the Floquet theory, also correctly applies to the current in this Andreev limit, as one has
\begin{equation}
	\overline{\left\langle I^A \right\rangle}_{q} = \sum_n P_n^A
	  I^A_{\text{DC}} \left( eV_{\text{DC}}+n\frac{\Omega}{2} \right),
\end{equation}
This may, however, have more to do with the linear in voltage behavior of the current in the DC biased case, than in the actual choice of weights to consider. Nevertheless, it is interesting to point out that both current and noise fit this modified Tien-Gordon picture in the Andreev regime.

It is interesting at this stage to compare the Tien-Gordon-type relations obtained in these two limiting regimes, Eqs.~\eqref{S_Normal_TG} and \eqref{S_Andreev_TG}. Indeed, keeping in mind that $P_n^A (q) = P_n (2q)$, one readily sees that despite their stark difference in physical origin, one being associated with QP-transfer, while the other is solely due to Andreev reflection, the two expressions show the same functional dependence, the main difference being the replacement $e \to 2e$. This effective doubling of the charge of the carriers when comparing the two limiting regimes was already apparent in Eq.~\eqref{eq:equivNA}. However, this goes one step further here, as it does not just contrast the results in the normal and the Andreev limit, but instead it compares how separately QP and Cooper pairs interact with the microwave radiation when the AC bias is applied, an interaction which thus bears striking similarities.

%
%

\subsubsection{General case}

Unfortunately, away from the previously considered limiting regimes, it is no longer possible to express the period-averaged noise as a simple weighted sum involving the noise of the DC-biased junction, and there is therefore no Tien-Gordon-type relation for the noise in the general case.

As we already noticed in our perturbative expansion, Eq.~\eqref{lambda2noiseTG}, at the lowest order in the tunneling constant $\lambda$, the $O \left( \lambda^2 \right)$ contribution to the noise does satisfy a Tien-Gordon-type relation, but this is mostly because it only involves one type of process, namely QP-transfer across the junction. Indeed, as soon as one includes next order contribution, the Tien-Gordon form breaks down irremediably, a feature which we attribute to the interference effects between the two main processes at play, QP-transfer and AR. While similar type of interference terms is also present in the pure DC case,~\cite{khlus1987a,muzykantskii1994a} as evidenced by the binomial factor $\tau_A(1-\tau_A)$, we believe they affect the noise in a sufficiently different way to prevent any kind of Tien-Gordon-type relation to hold anymore. Indeed, a major consequence of the periodic drive is to modify the description of the normal-metal into that of a superposition of energy-shifted Fermi seas.  This, in turn, modifies the behavior of the AR and NR correlations in a non-trivial way, as it now involves mixing between different Floquet channels, thus leading, as in Eq.~\eqref{lambda4noise}, to interference terms between two AR events (channels $n$ and $r$) and two NR events (channels $s-n$ and $s-r$).

One may wonder whether the noise in the general case, can be viewed as a combination of the Andreev and the quasiparticle contributions. Indeed, in the different but somewhat connected context of S-I-S junctions, it has been shown that the differential conductance can be written as a weighted sum of the supercurrent and quasiparticle current, which both satisfy a Tien-Gordon-type relation on their own, and separately describe the behavior of the system in limiting regimes~\cite{falci1991a,roychowdhury2015a}. As it turns out, this simple fitting fails in our case, and one can never write the noise as a weighted sum of the Andreev and the quasiparticle contributions of Eqs.~\eqref{S_Normal_TG} and \eqref{S_Andreev_TG}. This further strengthens our interpretation that the noise is not a simple superposition of the contributions from both carriers (QP and Cooper pairs) but rather a deep interconnection between the two which arises from the presence of the AC voltage and cannot be disentangled in the general situation.

\section{On-demand source of Cooper pairs}\label{OnDemand}

Building on our investigation of the driven $N-S$ junction and our understanding of the various regimes, an interesting problem is to inquire whether this superconductor hybrid device can be employed to achieve an on-demand source of Cooper pairs, a superconducting analog of the levitons in normal-metal devices~\cite{levitov1996a,dubois2013b}.

Naturally, such a source heavily relies on Andreev reflection, which thus requires us to operate as much as possible in the Andreev regime, ensuring that the drive frequency as well as the temperature are much smaller than the superconducting gap. Before analyzing the potential for the $N-S$ junction to be used as an on-demand source of Cooper pairs, let us first assess to what extent this Andreev regime is accessible experimentally. Borrowing from previous experiments on levitons in metals,~\cite{dubois2013b} one may estimate the typical orders of magnitude of the relevant parameters, with an electron temperature $\theta \approx \unit{10}{\milli\kelvin}$ and a drive frequency $f \approx \unit{5}{\giga\hertz}$ (note that this means applying voltages of the order $V_{\text{DC}} \approx\unit{10}{\micro\volt}$ to reach $q=0.5$). Choosing a junction involving niobium for the superconducting lead (with a typical gap $\Delta_\text{Nb}\approx \unit{1.55}{\milli\electronvolt} $) puts us well into the Andreev regime, as we have $\beta \Delta_\text{Nb} \approx 2000$ and $\frac{\Delta_\text{Nb}}{\Omega} \approx 100$. For a junction made out of aluminum ($\Delta_\text{Al} \approx \unit{0.17}{\milli\electronvolt}$), however, the situation is not as optimal, with  $\beta \Delta_\text{Al} \approx 200$ and $\frac{\Delta_\text{Al}}{\Omega} \approx 10$.

To achieve a reliable, controlled source of Cooper pairs, one needs to ensure two important properties: the average charge transmitted through the junction per period should be quantized, corresponding to an even number of electrons, and the excess noise should be vanishingly small so as to generate minimal excitation states of the $N-S$ junction.

Focusing on this regime, the average charge transmitted per period is defined as
\begin{equation}
	\langle Q \rangle = 2\pi\frac{\overline{\left\langle I^A \right\rangle}_{q}}{\Omega}=4 q e \tau_A\, ,
	\label{eq:avgQ}
\end{equation}
where we used the results of Eq.~\eqref{eq:IAndreev}. An ideal source of Cooper pairs would thus require to operate at perfect transmission and tune the bias voltage such that the injected charge $q$ is half-integer, leading to the emission of exactly $2q\in\mathbb{N}$ Cooper pairs per period into the superconductor. Quite remarkably, this result turns out to be independent of the type of drive considered.

Following our numerical results displayed in Fig.~\ref{Excess_noise_Andreev}, we can argue that only the half-quantized periodic Lorentzian drive is susceptible to show noise suppression. This can be further demonstrated by considering the expression for the excess noise in the Andreev regime, obtained from  Eq.~\eqref{Eq_Noise_andreev}, focusing on the zero temperature limit, where one has
\begin{align}
S_\text{exc}^A (q) \propto \sum_n P_n^A \left| 2 q + n \right| \left[ 1 - \text{Sgn} \left( 2 q + n \right)\right]
\end{align}
Since this is a sum of positive terms, the suppression of the excess noise requires every single term to vanish. While this is obviously the case for all terms $n > -2q$, for the other contributions to vanish, one must ensure that $P_n^A = 0$ for all $n \leq -2q$ which is only satisfied by a periodic Lorentzian drive with half-integer injected charge~\cite{dubois2013b,rech2017a}. 

From this, it follows that the ideal source of Cooper pairs corresponds to the $N-S$ junction in the Andreev regime driven by a periodic Lorentzian drive with half-integer injected charge $q$, and operating at zero temperature with perfect transmission. In practice, of course, none of these two latter conditions can be realistically met in an actual experiment. We now investigate what happens when we relax these constraints.

\begin{figure}[t]
	\centering
	\includegraphics[width=0.48\textwidth]{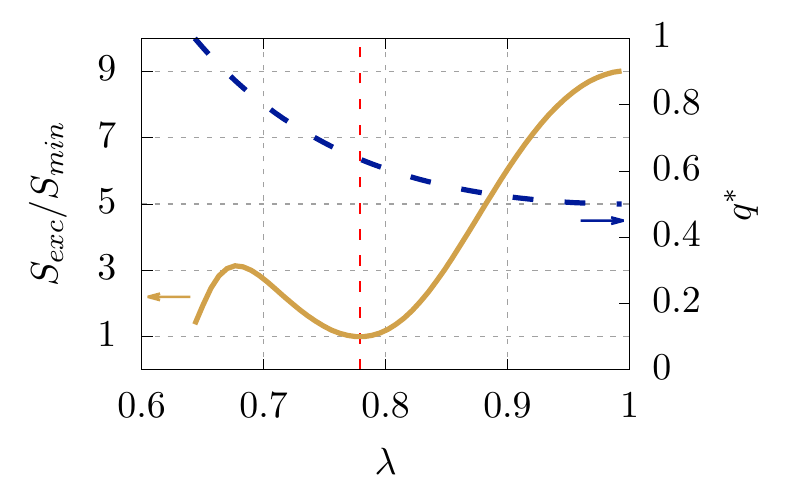}
	\caption{The excess noise in units of the minimal noise $S_\text{min}$ (full line, left axis) and the adjusted value of the injected charge per period (dotted line, right axis) as a function of the tunnel parameter $\lambda$ in the Andreev regime, $\Omega=10^{-2} \Delta$, at temperature $\theta = 5 \times 10^{-4} \Delta$. We restrict ourselves to values of $\lambda$ covering the interval $\tau_A \in [0.5 ; 1 ]$. The dotted red line corresponds to a transparency $\tau_A \simeq 0.79$ for which the excess noise is minimal.}
	\label{On_demand_CP}
\end{figure}

As it turns out, the average charge transferred is quite robust under variations of the electron temperature, so that the zero-temperature result of Eq.~\eqref{eq:avgQ} provides a very good estimate, even in the sub-optimal Andreev scenario of aluminum. This  linear in $\tau_A$ behavior suggests, however, that while working at finite temperature should hardly affect $\left\langle Q \right\rangle$, departing from perfect transmission has severe consequences, as the transmitted charge is no longer quantized. One way to circumvent this issue is to adjust the injected charge to a new value $q^*$ in order to reach the ideal value $\left\langle Q \right\rangle = 2 e$ for the source of Cooper pairs.

The excess noise mostly depends on the transmission through the overall prefactor, $\tau_A \left( 1 - \tau_A \right)$, typical of shot noise. Obviously, while this prefactor vanishes for perfect transmission, it is an unrealistic situation from the experimental standpoint. The dependence on temperature is a lot less trivial, and the excess noise is very much sensitive to a nonzero temperature $\theta$. Indeed, as one increases $\theta$, the results of Fig.~\ref{Excess_noise_Andreev} get modified leading to a first arch which instead of vanishing exactly at $q=0.5$, now reaches a nonzero local minimum for a slightly higher value of $q$. The value $S_\text{min}$ of this local minimum, which, in the Andreev regime, only depends on the drive frequency and temperature, is a good reference point to analyze the quality of the source.

Let us thus consider the $N-S$ junction and study how the excess noise varies for various values of the tunneling constant $\lambda$ when the voltage source is operated so as to maintain a quantized value $\left\langle Q \right\rangle = 2 e$ for the average transmitted charge. In Fig.~\ref{On_demand_CP}, we show the ratio $s = \frac{S_\text{exc}^A}{S_\text{min}}$ as a function of $\lambda$ for an injected charge $q^* =  \frac{1}{2 \tau_A}$ and the values of parameters mentioned earlier, $\theta = 5 \times 10^{-4} \Delta$ and $\Omega = 10^{-2} \Delta$ (which correspond to the situation of the Niobium junction). Quite remarkably, there is an interval of values of the tunneling constant (around $\lambda \simeq 0.78$, i.e., $\tau_A \simeq 0.79$) for which the excess noise is very close to the minimum allowed for this choice of temperature and drive frequency. It follows that there exists a set of experimentally reachable parameters (temperature, drive frequency, transparency) for which an actual reliable realization of this source of Andreev levitons can be envisioned, with an average transmitted charge $2 e$ and minimal excess noise. While it would be interesting to study in more details the properties of this source of Cooper pairs, this goes beyond the scope of this work.

\section{Conclusion}\label{Conclu}

In this paper, we studied the behavior of the current and noise in an $N-S$ junction subject to a periodic bias using a microscopic approach to all orders in the tunnel coupling. Our goal was to understand the detailed behavior of the junction and to state whether an equivalent of levitons could be envisioned in normal-metal-superconductor junctions, by studying the noise in excess compared to the DC biased junction. 

We found exact analytical results for both the average current and the zero-frequency period-averaged noise in the limit of large and small superconductor gap. An excess noise analysis was performed and we showed that, in the small-gap regime, for a Lorentzian train of pulses which bear a charge $e$ the junction displays minimal excitation states. On the other hand, in the infinite gap regime, the excess noise is suppressed for Lorentzian pulses injecting half-integer charge per period (dubbed Andreev levitons), which create an integer number of Cooper pairs in the superconductor. This analysis reveals an intriguing connection between these two limiting regimes where current and noise can be readily obtained by doubling the effective charge of the carriers and exchanging the transmission coefficients. 

When the gap is comparable to the frequency, the average current can be expressed analytically, and for low transparencies it displays strong kinks at the boundaries of the gap of each channel, surrounding integer values of the integer charge. These sharp discontinuities mark the switch from an AR-driven current to a QP-transfer-driven one. In this regime, the AC noise can be lesser than the DC one, showing that the usual correspondence between the excess noise and the number spurious electron hole pairs created by the drive does not hold anymore. Furthermore, the noise can no longer be expressed in a self-contained form, and we had to resort to a perturbative expansion in the tunneling constant, or to a numerical evaluation to all orders in $\lambda$, proving that the Floquet channel gap edges are responsible for the structures (cusps and minima) of the XN. 

Inspecting more closely our analytic derivations, we showed that the average current follows a Tien-Gordon behavior,~\cite{tien1963a} for the whole range of parameters. The same goes for the noise, which also satisfies Tien-Gordon-type relations, but only in the limiting regimes, dominated by QP-transfer or Andreev reflection. We could provide an interpretation of these results in terms of the leading physical processes also accounting for the interference effects expected between different Floquet channels. This Tien-Gordon formulation allowed us to uncover that not only the normal and Andreev regimes are related by an effective doubling of the charge, but also that the way QP and Cooper pairs interact with the microwave background is subject to the same kind of connection.

Finally, we studied the behavior of the noise and the charge transferred to the superconductor per period in order to characterize the controlled creation of Cooper pairs in the superconductor in the Andreev regime. As it turns out that, it is possible to find a regime of operation of the junction for which the average charge transferred is an even integer and the noise closely approaches its minimal value. Our simulations were performed at temperature, frequency, voltage and transparency already realized in previous experiments. Note that the present protocol differs from the usual way of designing such Cooper pair sources, typically with quantum pumps involving networks of superconducting islands creating Cooper pairs through Josephson physics,~\cite{erdman2019a} or ac-driven Coulomb blockade~\cite{geerligs1991a}. Our proposal is closer in spirit to the driven metallic junction~\cite{dubois2013b} or the mesoscopic capacitor~\cite{feve2007a}. This opens the way to using such a $N-S$ junction as a reliable, on-demand source of Cooper pairs, a fascinating perspective which requires more detailed investigation in the future. 

To summarize, in the limits of low and high frequencies compared to the superconducting gap, we extended the results of previous works~\cite{belzig2016a} to finite temperature. We provided a simple interpretation of the physics at play in terms of Floquet channels and Tien-Gordon like relations when applicable. Most importantly, we were able to quantify the full crossover regime of intermediate frequencies (comparable to the gap) and showed that the negative excess noise can be attributed to the fact that the density of states in the superconductor is not constant near the gap edges. Furthermore, we showed that the Tien-Gordon relations for the noise (established in the limit where the gap is larger than the drive frequency) break down in the intermediate regime because of interferences between different Floquet channels. The present theoretical results could in principle be probed experimentally in normal-metal/superconducting junctions where a microwave photon source irradiates the normal-metal side, in the same spirit as in the pioneering work of Ref.~\onlinecite{kozhevnikov2000a} which dealt with diffusive metals and (only) a sinusoidal drive. Finally, we proposed a scheme to build an on-demand source of Cooper pairs available with currently accessible technology and we stressed that such a source could be experimentally probed by a measure of the excess noise. In this respect, one may resort to modern hybrid nanowire/superconductor junctions with a QPC located at the interface in order to effectively control the transmission, and thus to implement the on-demand Cooper pair source.

\begin{acknowledgements}

The project leading to this publication has received funding from Excellence Initiative of Aix-Marseille University - A*MIDEX, a French “Investissements d’Avenir” program through the IPhU (AMX-19-IET-008) and AMUtech (AMX-19-IET-01X) institutes.
\end{acknowledgements}

\begin{appendix}

	\section{Double Fourier transform}\label{FourierApp}
	In this appendix we introduce the formalism for functions with a double time dependence. A two time periodic function $F$ can be written
	\begin{equation}
		\begin{aligned}
			F(t,t')=\sum_{nm}\int_{-\Omega/2}^{\Omega/2} & \frac{d\omega}{2\pi} e^{-i(\omega+n\Omega)t}\qquad \\
			                                             & \times e^{i(\omega+m\Omega)t'}F_{nm}(\omega)\, ,
		\end{aligned}
	\end{equation}
	with
	\begin{equation}
		\begin{aligned}
			F_{mn}(\omega)=\int_{-T/2}^{T/2} & \frac{\mathrm{d}\overline{t}}{T}\int_{-\infty}^{\infty}\mathrm{d}\tau e^{i(\omega+n\Omega)(\overline{t}+\frac{\tau}{2})} \\
			\times                           & e^{-i(\omega+m\Omega)(\overline{t}-\frac{\tau}{2})}F(\overline{t}+\tau/2,\overline{t}-\tau/2)\, .
		\end{aligned}
	\end{equation}
	The convolution product of two such functions is
	\begin{equation}
		\begin{aligned}
			(F\circ G)(t,t')=\sum_{m n} \int_{-\Omega/2}^{\Omega/2}\frac{d\omega}{2\pi} & e^{-i(\omega+n\Omega)t}\qquad \\
			\times e^{i(\omega+m\Omega)t'}                                              & (F\circ G)_{nm}(\omega)\, ,
		\end{aligned}
	\end{equation}
	where the harmonic components of the convolution product are $(F\circ G)_{nm}(\omega)=\sum_q F_{nq}(\omega)G_{qm}(\omega)$.

	\section{Floquet coefficients}\label{FloquetApp}
	A great variety of drives were explored by Vanevic \textit{et al.}~\cite{vanevic2008a} while here only three of them are exposed: the sinusoidal, square and Lorentzian ones as they display the three most different behaviors.
	\subsection{Sinusoidal}
	The voltage is
	\begin{equation}
		V(t) = V_{\text{DC}}(1-\cos{\Omega t})\, .
	\end{equation}
	This yields
	\begin{equation}
		p_{l} = J_l(-q)\, ,
	\end{equation}
	$J_l$ being the $l^{\text{th}}$ Bessel functions of the first kind.

	\subsection{Square}
	The bias is
	\begin{equation}
		V(t) = V_{\text{DC}} \left[ 1+\text{sgn}\left(\cos{\Omega t}\right) \right] \, .
	\end{equation}
	And the associated coefficient is
	\begin{equation}
		p_{l} = \frac{2}{\pi}\frac{q}{l^2-q^2}\sin \left[ {\frac{\pi}{2}(l - q)} \right] \, .
	\end{equation}

	\subsection{Lorentzian pulses}
	The train of Lorentzian pulses is defined by
	\begin{equation}
		V(t) = V_{\text{DC}} \left( \frac{1}{\pi} \sum_k \frac{\eta}{\eta^2 + (t/T - k)^2 }  \right)\, .
	\end{equation}
	with $\eta = W/T = W\Omega/2\pi$ the ratio between the width of the pulses and the period of the drive. This leads to a Fourier coefficient~\cite{rech2017a}
	\begin{equation}
		p_l	 = \int_{-1/2}^{1/2}du\, e^{2i\pi(l+q)u}\left(\frac{\sin[\pi(i\eta + u)]}{\sin[\pi(i\eta - u)]}\right)^{q}\, .
	\end{equation}

	\section{The bare Green functions}\label{BareGFApp}
	In full generality the bare Green functions in Nambu space are defined as
	\begin{equation}
		g^{\eta\eta'}(t,t')=-i\left\langle T_K\left(c(t_\eta)c^\dagger(t'_{\eta'})\right)\right\rangle\, .
	\end{equation}
	\subsection{The advanced an retarded Green functions}
	In the case of the BCS superconductor, the Fourier transform of the Green functions is given by~\cite{doniach1998a}
	\begin{equation}
		g_{\text{BCS}}^{r/a}(\omega)=\lim_{\delta\to0}\frac{\omega \openone +\Delta\sigma_x}{\sqrt{1-(\omega\pm i\delta)^2}}\, .
	\end{equation}
	The normal-metal corresponds to the zero gap limit of the BCS superconductor, i.e.,
	\begin{equation}
		g_{N}^{r/a}(\omega)=\mp i\openone \, .
	\end{equation}

	\subsection{The greater and lesser Green functions}
	The time-dependent voltage is defined as $V(t)=V_{\text{DC}} +V_{ac}(t)$ where the AC component averages to zero over one period, therefore, the energy entering the Fermi function can be written
	\begin{equation}
		\epsilon^\pm(\omega)=\omega\pm e V_{\text{DC}} \, ,
	\end{equation}
	the sign depending on the Nambu component, describing a hole or an electron.

	We want to compute the greater and lesser bare Green functions, defined~\cite{kamenev2011a} as $2g^{\pm\mp}=\mp g^r \pm g^a + g^k$ and $g^k=g^r(1-2F(\omega))-(1-2F(\omega))g^a$ with
	\begin{equation}
		F(\omega)=\begin{pmatrix}\frac{1}{1+\exp{\beta\epsilon^+(\omega)}} & 0                                         \\
               0                                         & \frac{1}{1+\exp{\beta\epsilon^-(\omega)}}
		\end{pmatrix}\, .
	\end{equation}
	The commutator $\left[g^{r/a},F(\omega)\right]$ vanishes in each lead, indeed, only the normal-metal is biased and its $r/a$ Green functions are diagonal in Nambu space, thus
	\begin{equation}
		g^{\pm\mp}(\omega)=i\text{Im}(g^{r})\left[\tanh\left(\frac{\omega+\sigma_zV_{0}}{2\theta}\right)\mp \openone \right]\, .
	\end{equation}

	\subsection{Floquet components}
	The bare Green functions only depend on the time difference $t-t'$. Thus, in the presence of a periodic drive, the harmonic components of the Fourier transform of any bare Green function in Nambu space can be written
	\begin{equation}
		g_{nm}^{\eta\eta'}(\omega)=g^{\eta\eta'}(\omega+n\Omega)\delta_{nm}\, .
	\end{equation}


	\section{Dyson equation}\label{DysonApp}
	We use Dyson equation with Keldysh formalism in order to write the dressed Green functions (including tunneling between the leads) in terms of the bare ones (which correspond to the isolated leads). The Retarded-Advanced-Keldysh (RAK) basis is defined as follows:~\cite{cuevas1999a}
	\begin{equation}
		\begin{aligned}
			G^{r} & =G^{-+}-G^{--}\, , \\
			G^{a} & =G^{+-}-G^{--}\, , \\
			G^{K} & =G^{++}+G^{--}\, .
		\end{aligned}
	\end{equation}

	The Dyson equation for the $2\times2$ Green function reads as (omitting convolution products)
	\begin{equation}
		\bold{G}=\bold{g}+\bold{g\Sigma G}\, .
	\end{equation}
	It is recursive and induces the following relations in Keldysh space:
	\begin{align}
		G^{\pm\mp} &= g^{\pm\mp}+g^{\pm\mp}\Sigma^{a}G^{a}+g^{r}\Sigma^{\pm\mp}G^{a}+g^{r}\Sigma^{r}G^{+-} \\
		G^{\pm\mp} &= (\openone +\Sigma^{r}G^{r})g^{\pm\mp}(\openone +\Sigma^{a}G^{a})-G^{r}\Sigma^{\pm\mp}G^{a}\, ,
	\end{align}
	The RAK components obey simpler Dyson equations
	\begin{equation}
		G^{r/a}=g^{r/a}+g^{r/a}\Sigma^{r/a}G^{r/a}\, ,
	\end{equation}
	where a summation over lead index is implicit in the notation. 
	
	
	After some algebra, one can show that
	\begin{widetext}
		\begin{gather}\label{DGF}
			\hat{G}_{RL}^{\pm\mp}=(\hat{\openone} -\lambda ^2\hat{g}_{R}^{r}\hat{\mathcal{P}}\hat{g}_{L}^{r}\hat{\mathcal{P}}^\dagger)^{-1}\big[\hat{g}_{R}^{\pm\mp}+ \hat{g}_{R}^{r} \hat{\mathcal{P}}\hat{g}_{L}^{\pm\mp}(\hat{\mathcal{P}}\hat{g}_{L}^{a})^{-1}\big](\hat{\openone} -\lambda^2 \hat{\mathcal{P}}\hat{g}_{L}^{a}\lambda \hat{\mathcal{P}}^\dagger \hat{g}_{R}^{a})^{-1}\lambda \hat{\mathcal{P}}\hat{g}_{L}^{a}\quad\text{and}\\[0.5cm]
			\hat{G}_{RR}^{\pm\mp}=(\hat{\openone} - \lambda^2 \hat{g}_{R}^{r}\hat{\mathcal{P}}\hat{g}_{L}^{r}\hat{\mathcal{P}}^\dagger)^{-1} (\hat{g}_{R}^{\pm\mp}+\lambda^2\hat{g}_{R}^{r}\hat{\mathcal{P}}\hat{g}_{L}^{\pm\mp}\hat{\mathcal{P}}^\dagger \hat{g}_{R}^{a}) (\hat{\openone} -\lambda^2 \hat{\mathcal{P}}\hat{g}_{L}^{a}\hat{\mathcal{P}}^\dagger \hat{g}_{R}^{a})^{-1}\, .
		\end{gather}
	\end{widetext}

	\onecolumngrid
	
	\section{Intermediate regime calculations}\label{CalculationApp}
	Dyson equations can be solved analytically as long as one of the leads is a normal-metal because $\hat{g}_N^{r/a}=\mp i\hat{\openone} $ so $\lambda^2 \hat{\mathcal{P}}^\dagger \hat{g}_N^{r/a} \hat{\mathcal{P}}=\mp\lambda^2i\hat{\openone} $ and the matrix to inverse is a tensor product between the identity in harmonics space and a $2\times2$ Nambu matrix.

	\subsection{Average current}
	With this in mind, one can compute the full Green function $G_{RL,nm}^{\pm\mp}$. Defining
	\begin{equation}
		\begin{aligned}
			\tilde{\omega}_n^{\pm}=\lim_{\delta\to0}\frac{\omega+n\Omega}{\sqrt{\Delta^2 - (\omega + n\Omega \pm i\delta)^2}}\, ,\qquad & \qquad\Delta_n^{\pm}=\lim_{\delta\to0}\frac{\Delta}{\sqrt{\Delta^2 - (\omega + n\Omega \pm i\delta)^2}} \, ,\\
			\xi_{n}^{\pm}=\frac{1}{1+\lambda^4\mp2i\lambda^2\omega_n^{\pm}}\, ,\qquad                                                   & \qquad\zeta_n^{\pm}=\left(\tanh(\omega+n\Omega)\mp1\right)                                              \, ,\\
			\Sigma_{LR,mn}^{r,a}=\lambda\begin{pmatrix}
				p_{n-m} & 0          \\
				0       & -p_{m-n}^*
			\end{pmatrix}\qquad                                                      & \qquad T_{nm}=\begin{pmatrix}
				\tanh(\frac{\omega+n\Omega+e V_{\text{DC}} }{2\theta}) & 0                                            \\
				0                                          & \tanh(\frac{\omega + n\Omega-e V_{\text{DC} }}{2\theta})
			\end{pmatrix}\delta_{nm}\, ,
		\end{aligned}
	\end{equation}
	the Green function entering the average current reads as 
	\begin{equation}
		\begin{aligned}
			G_{RL,nm}^{\pm\mp}=i\lambda\xi_n^+\xi_r^-\bigg\{ & \sigma_x\mathcal{P}_{nq}T_q\mathcal{P}_{qr}^\dagger \mathcal{P}_{rm}\Big[\Delta_n^++i\lambda^2\Delta_n^+\omega_r^-\Big]+\sigma_x\mathcal{P}_{nq}T_q\mathcal{P}_{qr}^\dagger \sigma_x\mathcal{P}_{rm}\Big[-i\lambda^2\Delta_n^+\Delta_r^-\Big]                                    \\
			+                                                & \mathcal{P}_{nq}T_q\mathcal{P}_{qr}^\dagger \mathcal{P}_{rm}\Big[\omega_n^++i\lambda^2+i\lambda^2\omega_n^+\omega_r^--\lambda^4\omega_r^-\Big]+\mathcal{P}_{nq}T_q\mathcal{P}_{qr}^\dagger \sigma_x\mathcal{P}_{rm}\Big[-i\lambda^2\Delta_r^-\omega_n^++\lambda^4\Delta_r^-\Big] \\
			+                                                & \mathcal{P}_{nm}\Big[\pm\omega_n^+-i\zeta_n^\pm\bar{\omega}_n\pm i\lambda^2\mp\lambda^4\omega_n^-+\zeta_n^\pm/2+\lambda^2(\zeta_n^\pm/2\pm i)(\omega_n^-\omega_n^+-\Delta_n^-\Delta_n^+)\Big]                                                                                    \\
			+                                                & \sigma_x\mathcal{P}_{nm}\Big[-i\zeta_n^\pm\bar{\Delta_n}+\lambda^2(\Delta_n^+\omega_n^--\Delta_n^-\omega_n^+)(\zeta_n^\pm/2\pm i) \pm(\Delta_n^++\lambda^4\Delta_n^-)\Big]\bigg\}\, ,
		\end{aligned}
	\end{equation}
	where we introduced $T_n$ such that $T_{nm} = T_n \delta_{nm}$. In order to compute the average current, one has to trace over Nambu and harmonics spaces, removing all the non diagonal terms, yielding
	\begin{align}
		\overline{\left\langle I \right\rangle}_{q}=& 2e\lambda^2\sum_{k} P_{k}(q) \int_{-\Omega/2}^{\Omega/2}\frac{\mathrm{d}\omega}{2\pi}\,\sum_{n} \xi_n^+\xi_n^-\text{Re}  \big[\lambda^2(\Delta_n^-\Delta_n^++  \omega_n^+\omega_n^-)-i\omega_n^+ + \lambda^2 +i\lambda^4\omega_n^-\big]    \nonumber   \\
		& \qquad \times \left[\tanh\left(\frac{\omega-(k+n)\Omega+eV_{\text{DC}}}{2\theta}\right) - \tanh\left(\frac{\omega+(k+n)\Omega-eV_{\text{DC}}}{2\theta}\right)\right] \nonumber \\
=& 2e\lambda^2\sum_{k} P_{k}(q) \int_{-\infty}^{\infty}\frac{\mathrm{d}\omega}{2\pi} {\cal I}(\omega) \left[\tanh\left(\frac{\omega - k \Omega+eV_{\text{DC}}}{2\theta}\right) - \tanh\left(\frac{\omega + k \Omega-eV_{\text{DC}}}{2\theta}\right)\right] \, ,
	\end{align}
with
\begin{equation}
{\cal I} (\omega)    = \left\{
\begin{aligned}
			\frac{2\lambda^2}{\left(1+\lambda^4 \right)^2 \left(1 - \frac{\omega^2}{\Delta^2} \right) + 4 \lambda^4 \frac{\omega^2}{\Delta^2}}                 & \qquad\text{if}\quad\left\lvert\omega\right\rvert<\Delta     \\
			\frac{1}{\left( 1 + \lambda^4 \right) \sqrt{1 - \frac{\Delta^2}{\omega^2}}+2 \lambda^2} & \qquad\text{if}\quad\left\lvert\omega\right\rvert>\Delta\, .
\end{aligned}\right.
\end{equation}

	\subsection{$\lambda^4$ noise}
	Here we compute the $\lambda^4$ contribution associated with PAN, solely due to AR.
	\paragraph{The first term of the noise.}
	We want to compute the first term of Eq.~\eqref{Noise_Fourier_global}. It can be written, after performing the trace over harmonics degrees of freedom, as
	\begin{equation}
		\begin{gathered}
			\sigma_z\Sigma_{LR,nq}G_{RL,qr}^{+-}\sigma_z\Sigma_{LR,rm}G_{RL,mn}^{-+}\\=-\lambda^4\sigma_z\mathcal{P}_{nq}^\dagger\bigg\{\sigma_x\mathcal{P}_{qr}T_r\Delta_q^++\mathcal{P}_{qr}T_q\omega_q^++\mathcal{P}_{qr}\Big[\omega_q^+-i\zeta_q^+\bar{\omega}_q\Big]+\sigma_x\mathcal{P}_{qr}\Big[-i\zeta_q^+\bar{\Delta}_q+\Delta_q^+\Big]\bigg\}\\
			\times\sigma_z\mathcal{P}_{rm}\bigg\{\sigma_x\mathcal{P}_{mn}T_n\Delta_m^++\mathcal{P}_{mn}T_n\omega_m^++\mathcal{P}_{mn}\Big[-\omega_m^+-i\zeta_m^-\bar{\omega}_m\Big]+\sigma_x\mathcal{P}_{mn}\Big[-i\zeta_m^-\bar{\Delta}_m-\Delta_m^+\Big]\bigg\}\, .
		\end{gathered}
	\end{equation}
	When looking at Andreev process in the gap only, both $\omega^{\pm}$ and $\Delta^{\pm}$ are real so $\bar{\omega}=\bar{\Delta}=0$, $\omega^+=\omega^-$ and $\Delta^-=\Delta^+$. The trace of this term therefore reduces to
	\begin{equation}
		\begin{gathered}
			\text{Tr}_{\text{N}}\Big[\sigma_z\Sigma_{LR,nq}G_{RL,qr}^{+-}\sigma_z\Sigma_{LR,rm}G_{RL,mn}^{-+}\Big]=\text{Tr}_{\text{N}}\Big[-\lambda^4\left(-\sigma_x\mathcal{P}_{qr}T_r\mathcal{P}_{rm}^\dagger\Delta_q+\mathcal{P}_{qr}T_r\mathcal{P}_{rm}^\dagger\omega_q+\delta_{qm}\omega_q-\delta_{qm}\sigma_x \Delta_q\right)\\
				\times\left(\sigma_x\mathcal{P}_{mn}T_n\mathcal{P}_{nq}^\dagger\Delta_m+\mathcal{P}_{mn}T_n\mathcal{P}_{nq}^\dagger\omega_m-\delta_{mq}\omega_m-\delta_{mq}\sigma_x\Delta_m\right)\Big]\, .
		\end{gathered}
	\end{equation}
	The noise is obtained by performing a Nambu trace so only the diagonal terms are kept, yielding
	\begin{equation}\label{NS_AC_Inter_Noise_O41}
		\begin{aligned}
			\text{Tr}_{\text{N}}\Big[\sigma_z\Sigma_{LR,nq}G_{RL,qr}^{+-} & \sigma_z\Sigma_{LR,rm}G_{RL,mn}^{-+}\Big]\\=
			\text{Tr}_{\text{N}}\Big[-                                    & 2\lambda^4\left(-\sigma_x\mathcal{P}_{qr}T_r\mathcal{P}_{rm}^\dagger\sigma_x\mathcal{P}_{mn}T_n\mathcal{P}_{nq}^\dagger\Delta_m\Delta_q + \mathcal{P}_{qr}T_r\mathcal{P}_{rm}^\dagger \mathcal{P}_{mn}T_n\mathcal{P}_{nq}^\dagger\omega_m\omega_q+\openone (\Delta_m^2-\omega_m^2)\right)\Big]\, .
		\end{aligned}
	\end{equation}
	\paragraph{The second term.}
	We want to compute the second term of Eq.~\eqref{Noise_Fourier_global}. We start by computing (note that some permutations have been performed as one is only interested in the trace)
	\begin{equation}
		\begin{gathered}
			\lambda^2 \hat{\sigma}_z \hat{\mathcal{P}} \hat{G}_{LL}^{\pm\mp} \hat{\sigma}_z \hat{\mathcal{P}}^\dagger \hat{G}_{RR}^{\mp\pm}\\=
			\lambda^2\hat{\sigma}_z(\hat{\openone} -\lambda^2 \hat{g}_{R}^{r} \hat{\mathcal{P}} \hat{g}_{L}^{r} \hat{\mathcal{P}}^\dagger)^{-1} \hat{\mathcal{P}}\big(\hat{g}_{L}^{\pm\mp}+ \lambda^2 \hat{g}_{L}^{r} \hat{\mathcal{P}}^\dagger \hat{g}_{R}^{\pm\mp} \hat{\mathcal{P}} \hat{g}_{L}^{a}\big) \hat{\mathcal{P}}^\dagger(\hat{\openone} -\lambda^2 \hat{\mathcal{P}} \hat{g}_{L}^{a} \hat{\mathcal{P}}^\dagger \hat{g}_{R}^{a})^{-1}\hat{\mathcal{P}}\hat{\sigma}_z \hat{\mathcal{P}}^\dagger\\
			\times (\hat{\openone} -\lambda^2 \hat{g}_{R}^{r} \hat{\mathcal{P}} \hat{g}_{L}^{r} \hat{\mathcal{P}}^\dagger)^{-1}\big(\hat{g}_{R}^{\mp\pm}+ \lambda^2 \hat{g}_{R}^{r} \hat{\mathcal{P}} \hat{g}_{L}^{\mp\pm} \hat{\mathcal{P}}^\dagger \hat{g}_{R}^{a}\big)(\hat{\openone} -\lambda^2 \hat{\mathcal{P}} \hat{g}_{L}^{a} \hat{\mathcal{P}}^\dagger \hat{g}_{R}^{a})^{-1}\, .
		\end{gathered}
	\end{equation}
	Some simplifications can be performed, first $\hat{g}_L$ commutes in Nambu and harmonics space with $ \hat{\mathcal{P}}^\dagger$ so the second parentheses can be simplified. Furthermore, the inverse of $(\hat{\openone} -\lambda^2 \hat{\mathcal{P}} \hat{g}_{L}^{a} \hat{\mathcal{P}}^\dagger \hat{g}_{R}^{a})^{-1}$ is diagonal in harmonic space, the same goes for $g_L$. As a result, the frequencies at which this term is evaluated will be that of $g_{R}^{\pm\mp}$ and therefore it vanishes when looking at in-gap AR. The next step is to keep only $\lambda^4$ terms. As
	\begin{equation}
		(\hat{\openone} -\lambda \hat{\mathcal{P}} \hat{g}_{L}^{a}\lambda \hat{\mathcal{P}}^\dagger \hat{g}_{R}^{a})^{-1}_{nm} \sigma_z (\hat{\openone} -\hat{g}_{R}^{r}\lambda \hat{\mathcal{P}} \hat{g}_{L}^{r}\lambda \hat{\mathcal{P}}^\dagger)^{-1}_{mq}= \sigma_z(\xi_n^+\xi_n^-)^2 \delta_{nq} = \sigma_z \delta_{nq}+o(\lambda^4)\, ,
	\end{equation}
	the total term to evaluate becomes
	\begin{equation}
		\left[ \lambda \hat{\mathcal{P}}\hat{g}_{L}^{\pm\mp}\lambda \hat{\mathcal{P}}^\dagger\hat{\sigma}_z \hat{g}_{R}^{r}\lambda \hat{\mathcal{P}}^{r} \hat{g}_{L}^{\mp\pm}\lambda \hat{\mathcal{P}}^\dagger \hat{g}_{R}^{a}\hat{\sigma}_z \right]_{nn}=-\lambda^4\mathcal{P}_{nq}(T_q\mp\openone )\mathcal{P}_{qr}^\dagger\sigma_z(\omega_r\openone +\Delta_r\sigma_x)\mathcal{P}_{rs}(T_s\pm\openone )\mathcal{P}_{sn}^\dagger(\omega_n\openone +\Delta_n\sigma_x)\sigma_z\, .
	\end{equation}
	As the $\mp$ and the $\pm$ results are summed in the noise so only half of the terms will count and the sum yields
	\begin{equation}
		\begin{gathered}
			\left[-\hat{\sigma}_z\lambda \hat{\mathcal{P}}\hat{G}_{LL}^{+-}\hat{\sigma}_z\lambda \hat{\mathcal{P}}^\dagger \hat{G}_{RR}^{-+}-\hat{\sigma}_z\lambda \hat{\mathcal{P}}^\dagger \hat{G}_{RR}^{-+}\hat{\sigma}_z\lambda \hat{\mathcal{P}}\hat{G}_{LL}^{+-} \right]_{nn}\\=2\lambda^4\mathcal{P}_{nq}\Big[T_q\mathcal{P}_{qr}^\dagger\sigma_z(\omega_r\openone +\Delta_r\sigma_x)\mathcal{P}_{rs}T_s\mathcal{P}_{sn}^\dagger(\omega_n\openone +\Delta_n\sigma_x)-   \mathcal{P}_{qr}^\dagger\sigma_z(\omega_r\openone +\Delta_r\sigma_x)\mathcal{P}_{rs}\mathcal{P}_{sn}^\dagger(\omega_n\openone +\Delta_n\sigma_x)\Big]\sigma_z\, .
		\end{gathered}
	\end{equation}
	The term contributing to the trace therefore reduces to
	\begin{equation}\label{NS_AC_Inter_Noise_O42}
		\text{Tr}_{\text{NH}}\Big[\lambda^2\hat{\sigma}_z \hat{\mathcal{P}}\hat{G}_{LL}^{\pm\mp}\hat{\sigma}_z \hat{\mathcal{P}}^\dagger \hat{G}_{RR}^{\mp\pm}\Big] =2\lambda^4 \sum_n \text{Tr}_{\text{N}}\Big[\mathcal{P}_{nq}T_q\mathcal{P}_{qr}^\dagger \mathcal{P}_{rs}T_s\mathcal{P}_{sn}^\dagger\omega_n\omega_r+\sigma_x\mathcal{P}_{nq}T_q\mathcal{P}_{qr}^\dagger\sigma_x \mathcal{P}_{rs}T_s\mathcal{P}_{sn}^\dagger\Delta_n\Delta_r - \openone (\omega_n^2 + \Delta_n^2)\Big]\, .
	\end{equation}

	\paragraph{Sum of the terms.}
	Summing both terms yields
	\begin{equation}
		\overline{\left\langle S \right\rangle}_{q}=-8e^2\lambda^4\int_{-\Omega/2}^{\Omega/2}\frac{\mathrm{d}\omega}{2\pi}\sum_{nsr=-\infty}^{+\infty}\text{Tr}_{\text{N}}\left[\sigma_x\mathcal{P}_{nq}T_q\mathcal{P}_{qr}^\dagger\sigma_x \mathcal{P}_{rs}T_s\mathcal{P}_{sn}^\dagger\Delta_n\Delta_r - \Delta_n^2\openone \right]\, .
	\end{equation}
	After performing the Nambu trace, redefining the indices and performing one harmonics sum one is left with Eq.~\eqref{lambda4noise}.
\end{appendix}
\twocolumngrid

\end{document}